\documentclass[numberedappendix]{emulateapj} 

\usepackage {epsf}
\usepackage{epstopdf}
\usepackage {lscape, graphicx}

\gdef\1054{MS\,1054--03}

\def\farcs{\hbox{$.\!\!^{\prime\prime}$}}
\def\simgeq{{\raise.0ex\hbox{$\mathchar"013E$}\mkern-14mu\lower1.2ex\hbox{$\mathchar"0218$}}} 

\begin {document}

\title {Galaxy Structure and Mode of Star Formation in the SFR-Mass Plane from $z \sim 2.5$ to $z \sim 0.1$}

\author{Stijn Wuyts\altaffilmark{1}, Natascha M. F\"{o}rster Schreiber\altaffilmark{1},
Arjen van der Wel\altaffilmark{2}, Benjamin Magnelli\altaffilmark{1},
Yicheng Guo\altaffilmark{3}, Reinhard Genzel\altaffilmark{1},
Dieter Lutz\altaffilmark{1},
Herv\'{e} Aussel\altaffilmark{4},
Guillermo Barro\altaffilmark{5},
Stefano Berta\altaffilmark{1},
Antonio Cava\altaffilmark{6}, 
Javier Graci\'{a}-Carpio\altaffilmark{1},
Nimish P. Hathi\altaffilmark{7},
Kuang-Han Huang\altaffilmark{8},
Dale D. Kocevski\altaffilmark{5},
Anton M. Koekemoer\altaffilmark{9}, 
Kyoung-Soo Lee\altaffilmark{10},
Emeric Le Floc'h\altaffilmark{4}, 
Elizabeth J. McGrath\altaffilmark{5},
Raanan Nordon\altaffilmark{1}, Paola Popesso\altaffilmark{1},
Francesca Pozzi\altaffilmark{11}, Laurie Riguccini\altaffilmark{4},
Giulia Rodighiero\altaffilmark{12}, Amelie Saintonge\altaffilmark{1},
Linda Tacconi\altaffilmark{1}}

\altaffiltext{1}{Max-Planck-Institut f\"{u}r extraterrestrische Physik, Giessenbachstrasse, D-85748 Garching, Germany}
\altaffiltext{2}{Max-Planck-Institut f\"{u}r Astronomie, K\"{o}nigstuhl 17, D-69117 Heidelberg, Germany}
\altaffiltext{3}{Astronomy Department, University of Massachusetts, 710 N. Pleasant Street, Amherst MA01003, USA}
\altaffiltext{4}{Laboratoire AIM, CEA/DSM-CNRS-Universit\'{e} Paris Diderot, IRFU/Service d'Astrophysique, B\^{a}t. 709, CEA-Saclay, 91191 Gif-sur-Yvette Cedex, France}
\altaffiltext{5}{UCO/Lick Observatory, Department of Astronomy and Astrophysics, University of California, Santa Cruz CA 95064, USA}
\altaffiltext{6}{Departamento de Astrof\'{\i}sica, Facultad de CC. F\'{\i}sicas, Universidad Complutense de Madrid, E-28040 Madrid, Spain}
\altaffiltext{7}{Observatories of the Carnegie Institution of Washington, Pasadena CA 91101, USA}
\altaffiltext{8}{Johns Hopkins University, 3400 N. Charles Street, Baltimore MD 21218, USA}
\altaffiltext{9}{Space Telescope Science Institute, 3700 San Martin Drive, Baltimore MD 21218, USA}
\altaffiltext{10}{Yale Center for Astronomy and Astrophysics, Department of Physics, Yale University, New Haven CT 06520, USA}
\altaffiltext{11}{Dipartimento di Astronomia, Universit\`{a} di Bologna, via Ranzani 1, 40127 Bologna, Italy}
\altaffiltext{12}{Dipartimento di Astronomia, Universit\`{a} di Padova, Vicolo dell'Osservatorio 3, 35122 Padova, Italy}

\begin{abstract}

We analyze the dependence of galaxy structure (size and Sersic index) and mode of star formation ($\Sigma_{SFR}$ and $SFR_{IR}/SFR_{UV}$) on the position of galaxies in the SFR versus Mass diagram.  Our sample comprises roughly 640000 galaxies at $z \sim 0.1$, 130000 galaxies at $z \sim 1$, and 36000 galaxies at $z \sim 2$.  Structural measurements for all but the $z \sim 0.1$ galaxies are based on HST imaging, and SFRs are derived using a Herschel-calibrated ladder of SFR indicators.  We find that a correlation between the structure and stellar population of galaxies (i.e., a 'Hubble sequence') is already in place since at least $z \sim 2.5$.  At all epochs, typical star-forming galaxies on the main sequence are well approximated by exponential disks, while the profiles of quiescent galaxies are better described by de Vaucouleurs profiles.  In the upper envelope of the main sequence, the relation between the SFR and Sersic index reverses, suggesting a rapid build-up of the central mass concentration in these starbursting outliers.  We observe quiescent, moderately and highly star-forming systems to co-exist over an order of magnitude or more in stellar mass.  At each mass and redshift, galaxies on the main sequence have the largest size.  The rate of size growth correlates with specific SFR, and so does $\Sigma_{SFR}$ at each redshift.  A simple model using an empirically determined SF law and metallicity scaling, in combination with an assumed geometry for dust and stars is able to relate the observed $\Sigma_{SFR}$ and $SFR_{IR}/SFR_{UV}$, provided a more patchy dust geometry is assumed for high-redshift galaxies.

\end{abstract}


\section {Introduction}
\label{intro.sec}

Deep multi-wavelength lookback surveys carried out over the past five years have improved our understanding of galaxy evolution in the young universe tremendously.  At least over the past 10 Gyr of lookback time, a correlation between the rate of star formation and the amount of assembled stellar mass in star-forming galaxies (SFGs), dubbed the 'main sequence of star formation' (MS), has been observed (Noeske et al. 2007; Elbaz et al. 2007; Daddi et al. 2007).  While the scatter around the MS is not observed to evolve strongly with redshift, the zeropoint does, in the sense that high-redshift SFGs form stars at a higher rate than similar mass galaxies today. 

Complementing the observational results on the locus of SFGs in SFR-Mass space, direct measurements of the CO molecular line emission from galaxies residing on the MS have revealed molecular gas mass fractions as high as 35 (45) \% at $z \sim 1\ (2)$ (Tacconi et al. 2010).  Linking surface densities of gas and star formation, Genzel et al. (2010) recently argued that high-redshift SFGs follow the same star formation law as local SFGs, unlike their merging counterparts that have shorter depletion timescales and follow a Kennicutt-Schmidt relation with a higher zeropoint (see also Daddi et al. 2010).

 From the observed zeropoint evolution of the MS, and tightness of the relation, a picture has emerged in which the bulk of SFGs are forming stars gradually over timescales that are long (one to two orders of magnitude larger) relative to their dynamical times (Genzel et al. 2010; Wuyts et al. 2011).  This is only possible by a continuous replenishment of their gas reservoirs.  Cold, filamentary streams of gas have been proposed on the basis of cosmological simulations as an efficient mechanism to penetrate the galaxies' surrounding halos, and deposit new (or recycled) fuel for star formation (Keres et al. 2005, 2009; Dekel \& Birnboim 2006; Dekel et al. 2009).  Evidence from kinematics strongly supports such a continuous (i.e., non-major merger) triggering and maintenance of star formation for the bulk of SFGs (Genzel et al. 2008; F\"{o}rster Schreiber et al. 2009; Shapiro et al. 2008; F\"{o}rster Schreiber et al. 2011a).

In such a scenario, a secular mode of star formation dominates the cosmic SFR history, and the scatter around the MS can, aside from a contribution by measurement uncertainties, be attributed to burstiness induced by occasional (minor) merger events.  While the excess star formation due to mergers may only amount to 5-10\% (Hopkins et al. 2010; Rodighiero et al. 2011), they may play a crucial role in shaping the diversity of galaxies observed both at low and high redshift.  The merger paradigm (Sanders et al. 1988; Hopkins et al. 2006) links a wide variety of galaxy types in an evolutionary sequence in which the collision between SFGs creates tidal torques that channel gas to the center where it triggers a nuclear starburst (e.g., Barnes \& Hernquist 1991, 1996).  When the starburst is accompanied by accretion onto the central supermassive black hole(s) (SMBHs), this produces AGN emission, and naturally accounts for the origin of a correlation between SMBH and bulge mass (Di Matteo et al. 2005).  Finally, a combination of gas exhaustion, supernova and AGN feedback leads to a rapid decline in the SFR, leaving a post-starburst, red and dead galaxy as remnant (Wuyts et al. 2010; Snyder et al. 2011).

Indeed, such classes of galaxies offset from the MS have also been identified observationally out to $z \sim 2.5$ and even beyond: towards the high-SFR end with for instance bright sub-millimeter galaxies (SMGs, Smail et al. 1997; Barger et al. 1998; Hughes et al. 1998; see Blain et al. 2002 for a review), and towards the low-SFR end (Daddi et al. 2005; Kriek et al. 2006; van Dokkum et al. 2011).  Observations of their structural properties, albeit often limited to small samples, could lend credence to their merger connection.  Several of the SMGs imaged in CO show two-component morphologies and/or highly disturbed velocity fields, implying a major merger nature (Tacconi et al. 2008).  Massive ($\sim 10^{11}\ M_{\sun}$) quiescent galaxies at high redshift are increasingly more compact relative to similar mass galaxies today ($r_e / r_{e,\ z=0} \sim (1+z)^{-1.1  \pm 0.2}$) (e.g., van der Wel et al. 2008; van Dokkum et al. 2008, but see Mancini et al. 2010 for a contrasting view), in agreement with expectations from binary merger scenarios in which the gas content of the progenitors increases with redshift (Khochfar \& Silk 2006; Hopkins et al. 2009; Wuyts et al. 2010).  Due to the highly dissipational formation process, Wuyts et al. (2010) predict a larger degree of rotation in the remnants than in lower redshift early-type galaxies.  Observational evidence for high-redshift compact quiescent disks (McGrath et al. 2008; van der Wel et al. 2011) could possibly reflect this, or alternatively may signal other quenching mechanisms at play.

With the high-resolution optical (ACS) and near-infrared (WFC3) cameras onboard HST, detailed structural measurements can now be obtained for ever larger samples, bringing statistical significance also for the more extreme populations.  In particular, the ongoing Cosmic Assembly Near-infrared Deep Extragalactic Legacy Survey (CANDELS, Grogin et al. 2011; Koekemoer et al. 2011) is collecting deep rest-frame optical (and rest-UV) space-based imaging over unprecedented areas.  At the same time, the PACS Evolutionary Probe (PEP, Lutz et al. 2011) opens a window on the far-infrared emission of large samples of SFGs, and allows the calibration of other SFR indicators for use in fields that lack such (deep) far-IR data (Wuyts et al. 2011).

In this paper, we exploit high-resolution imaging and a cross-calibrated 'ladder of SFR indicators' to study the structure of galaxies and their mode of star formation, as function of their position in the SFR-Mass diagram.  We discuss how properties such as size, surface brightness profile shape, SFR surface density, and $IR/UV$ ratios vary along and across the MS, and how these trends evolve with redshift (from $z  \sim 2.5$ till today).  To this end, we assembled samples of unprecedented size, with 639924 galaxies at $z \sim 0.1$, 132328 galaxies at $z \sim 1$, and 35649 galaxies at $z \sim 2$.  The data quality and sample size allow us to extend previous studies of the relation between structure and stellar populations in the nearby (Kauffmann et al. 2003; Schiminovich et al. 2007; Bell 2008) and intermediate-redshift (Scarlata et al. 2007; Maier et al. 2009) universe, and to build on the ground-based explorations of these early epochs (Franx et al. 2008; Williams et al. 2010).

The paper is structured as follows.  We present an overview of the fields and data used, the methods to compute derived products, and the final sample in Section\ \ref{obs_method_sample.sec}.  Readers who wish to skip directly to our results, should turn to Sections\ \ref{structure.sec} and\ \ref{mode.sec}, where we analyze galaxy structure and the mode of star formation in the SFR-Mass diagram respectively.  Section\ \ref{discussion.sec} discusses the observational results in a physical context.  Finally, we summarize our conclusions in Section\ \ref{summary.sec}.

Throughout this paper, we quote magnitudes in the AB system, assume a Chabrier (2003) inital mass function (IMF), and
adopt the following cosmological parameters: $(\Omega _M, \Omega
_{\Lambda}, h) = (0.3, 0.7, 0.7)$.

\section{Observations, Methods and Sample}
\label{obs_method_sample.sec}

In our analysis, we divide the galaxies in three redshift bins: $0.02 < z < 0.2$, $0.5 < z < 1.5$, and $1.5 <  z < 2.5$, which we will in short refer to as $z \sim 0.1$, $z \sim 1$, and $z \sim 2$.  The $z \sim 0.1$ sample is extracted from the Sloan Digital Sky Survey (SDSS), as detailed in Section\ \ref{SDSS_GALEX.sec}.  The higher redshift samples are extracted from four deep fields that host some of the richest multi-wavelength data sets available to date: COSMOS, UDS, GOODS-South, and GOODS-North.  We describe the data available for them in Sections\ \ref{COSMOS.sec} to \ref{GOODSN.sec}, and summarize the methods applied to compute derived properties such as photometric redshifts, stellar masses, SFRs and structural parameters in Section\ \ref{derived_prop.sec}.

\subsection{Fields and Data}
\label{fields_data.sec}

\subsubsection{SDSS + GALEX}
\label{SDSS_GALEX.sec}

SDSS offers a robust local universe anchor for our study of galaxy properties in the SFR-Mass diagram, and its evolution with redshift.  We use the spectroscopic SDSS sample (DR7, Abazajian et al. 2009) in overlap between the MPA-JHU and NY-VAGC value-added galaxy catalogs.  SFRs and stellar masses were taken from the MPA-JHU compilation\footnote{http://www.mpa-garching.mpg.de/SDSS/DR7}.  Briefly, the stellar masses were estimated  by fitting Bruzual \& Charlot (2003, hereafter BC03) stellar population synthesis models with a wide range of star formation histories (SFHs) to the broad-band photometry (Salim et al. 2007).  Unless explicitly stated otherwise, we use total, dust- and aperture-corrected SFRs.  SFRs within the SDSS spectroscopic fiber were primarily based on H$\alpha$, using the calibration of Brinchmann et al. (2004), which includes an extinction correction based on the Balmer decrement.  Aperture corrections to total use fits to the photometry of the outer regions of the galaxies, as described by Salim et al. (2007).  We use the SDSS (DR7) - GALEX (GR5) matched catalog from Bianchi et al. (2011) to compute the unobscured part of the star formation $SFR_{UV}$.  Sizes and profile shapes were measured by the NY-VAGC team\footnote{http://sdss.physics.nyu.edu/vagc} by means of Sersic fits to the galaxy surface brightness profiles (Blanton et al. 2003, 2005).  By default, we adopt the structural parameters measured in the $g$ band, but we will discuss the dependence on wavelength where appropriate.

\subsubsection{COSMOS}
\label{COSMOS.sec}

We make use of public multi-wavelength photometry from Ilbert et al. (2009) and Gabasch et al. (2008) over an effective area of 1.48 deg$^2$ in the COSMOS HST field (Scoville et al. 2007; Koekemoer et al. 2007).  Our working catalog is cut at $i < 25$ to guarantee sufficient signal-to-noise to allow for a reliable determination of photometric redshifts and other derived properties.  A total of 36 medium and broad bands sample the SED from GALEX to IRAC wavelengths.  At infrared wavelengths, PACS imaging from PEP, extracted using 24 $\mu$m sources as prior, reaches depths of 5.7 and 11.9 mJy (3$\sigma$) at 100 $\mu$m and 160 $\mu$m respectively.  MIPS 24 $\mu$m imaging in COSMOS (Sanders et al. 2007; Le Floc'h et al. 2009) is used down to 60 $\mu$Jy.  An X-ray catalog based on XMM-Newton observations is released by Cappelluti et al. (2009).  High-resolution $I_{814}$-band imaging with HST/ACS ($I_{814} < 27,\ 5\sigma$) spanning the entire field, and probing the rest-frame $B$ and $2700\AA$ band at $z \sim 1$ and $z \sim 2$ respectively, forms the foundation for structural measurements in COSMOS.

\subsubsection{UDS}
\label{UDS.sec}

A deep WFC3 $H_{160}$-selected catalog ($H_{160} < 26.7$, 5$\sigma$) with consistent photometry in 16 bands from $B$ to 8 $\mu$m was constructed on the basis of CANDELS, SEDS (PI G. Fazio), SpUDS (PI J. Dunlop), and ancillary ground-based data in the $\sim 200$ arcmin$^2$ CANDELS-covered UDS area (Guo et al. in prep).  MIPS 24 $\mu$m imaging to a depth of 30 $\mu$Jy (3$\sigma$) was carried out as part of the SpUDS survey.  Ueda et al. (2008) presented the X-ray source catalog of the Subaru/XMM-Newton deep survey, in which the CANDELS/UDS area is embedded.  We performed the structural measurements on the WFC3 $H_{160}$ mosaic, which with 4/3 orbits per pointing forms part of CANDELS-Wide.  The $H_{160}$ band probes the rest-frame $I$ and $V$ band at $z \sim 1$ and $z \sim 2$ respectively.

\subsubsection{GOODS-South}
\label{GOODSS.sec}

Following similar procedures as for UDS, Guo et al. (in prep) constructed a deep WFC3 $H_{160}$-selected catalog ($H_{160} < 27$, 5$\sigma$) for the Early Release Science (ERS, PI O'Connell) and CANDELS-Deep area in GOODS-South.  The catalog contains consistent photometry in 14 passbands from $U$ to 8 $\mu$m.  Spanning the same wavelength range, we use FIREWORKS ($K_s < 24.3,\ 5\sigma$, Wuyts et al. 2008) as supporting multi-wavelength catalog to exploit CANDELS-Wide data in the bottom 20\% of the 148 sq. arcmin GOODS-South field.  As part of the PEP survey (Lutz et al. 2011), PACS photometry was obtained to a 5$\sigma$ depth of 1.8 mJy, 1.9 mJy and 3.3 mJy at 70 $\mu$m, 100 $\mu$m and 160 $\mu$m respectively, using the position of 24 $\mu$m sources as prior.  The 24 $\mu$m imaging itself (30 $\mu$Jy, 5$\sigma$) was taken from Magnelli et al. (2009).  Where appropriate, we select X-ray sources from the Chandra 2 Ms source catalog by Luo et al. (2008).  We performed the structural measurements on WFC3 $H_{160}$-band imaging drizzled to a $0\farcs06 / pixel$ scale, obtained as part of the ERS and CANDELS programs.  The WFC3/IR observations of the ERS were obtained in F098M, F125W and F160W for a total depth of 2 orbits per pointing in each filter (Windhorst et al. 2011) and the mosaics used here were produced using MultiDrizzle (Koekemoer et al. 2002) following the approach outlined in Koekemoer et al. (2011).  The 4-epoch CANDELS-Deep area is similar in depth, while CANDELS-Wide received half the integration time.

\subsubsection{GOODS-North}
\label{GOODSN.sec}

We derived photometric redshifts and stellar masses for galaxies in GOODS-North based on a $z+K$-selected catalog with photometry in 16 bands from GALEX to IRAC wavelengths (Berta et al. 2010).  Additional longer wavelength information at 24 $\mu$m (30 $\mu$Jy, 5$\sigma$), 100 $\mu$m (5.1 mJy, 5$\sigma$), and 160 $\mu$m (8.7 mJy, 5$\sigma$) was acquired as part of the GOODS and PEP surveys.  Where appropriate, we select X-ray sources from the Chandra 2 Ms source catalog by Alexander et al. (2003).  Structural measurements in GOODS-North are based on the GOODS ACS $z_{850}$-band imaging (Giavalisco et al. 2004), corresponding to the rest-frame $B$ and $3000\AA$ band at $z \sim 1$ and $z \sim 2$ respectively..

\subsection{Derived Galaxy Properties}
\label{derived_prop.sec}

For all deep fields, we use identical procedures to derive photometric redshifts, stellar masses, SFRs, and structural parameters, hence optimizing the consistency of the combined data set.

\subsubsection{Photometric Redshifts}

We determined photometric redshifts by fitting a superposition of 6 galaxy templates to the observed broad-band SEDs using the photometric redshift code EAZY (Brammer et al. 2008).  For the 6955 out of 132328 (5.3\%) of galaxies at $0.5 < z < 1.5$, and 497 out of 35649 (1.4\%) of galaxies at $1.5 < z < 2.5$ that have a spectroscopic redshift from one of the many spectroscopic campaigns in the above four fields (see, e.g., Vanzella et al. 2008; Lilly et al. 2009; Barger et al. 2008), we adopt the $z_{spec}$ as redshift in our analysis.

Comparing the photometric redshifts $z_{phot}$ and spectroscopic redshifts $z_{spec}$ of spectroscopically confirmed galaxies, we find a median and scatter (normalized median absolute deviation) in $\Delta z / (1+z)$ of (-0.001; 0.015) in the $z \sim 1$ and (-0.007; 0.052) in the $z \sim 2$ redshift bin.  Naturally, these uncertainties only apply to galaxies with similarly bright magnitudes as represented in the spectroscopic sample based on which the scatter was computed.  E.g., the public zCOSMOS-bright spectroscopic survey (Lilly et al. 2009) reaches down to $i = 22.5$.  For this spectroscopic sample in COSMOS, we measure a scatter in $\Delta z / (1+z)$ of 0.010.  As detailed in Appendix B.3, the estimated uncertainty increases to $\sigma_{NMAD} (\Delta z / (1+z)) = 0.048$ for COSMOS galaxies with $24 < i < 25$, which account for 40\% of the galaxies in that field, and 30\% of our entire sample.  The estimated uncertainty is consistent with earlier work by Ilbert et al. (2009).  For the other deep fields as well, the increase in the photometric redshift uncertainty with respect to the spectroscopic sample amounts to a factor of $\sim 5$ for the faintest galaxies that enter our analysis.  An extensive analysis in Appendix B.3 shows that our results are robust against photometric redshift uncertainties, even when accounting for the fact that the $z_{phot}$ quality of the entire sample is likely lower than that of the (typically brighter) spectroscopically confirmed subsample.

\subsubsection{Stellar Masses}
\label{SEDmodeling.sec}

We used FAST (Kriek et al. 2009a) to fit BC03 models with exponentially declining SFHs to the $\lambda_{obs} \leq 8\ \mu$m broad-band SEDs of the galaxies.  Following the recipe presented by Wuyts et al. (2011), we impose a minimum e-folding time of 300 Myr.  The time since the onset of star formation was allowed to vary between 50 Myr and the age of the universe at the observed redshift.  Our choice of SFHs guarantees consistency with most of the literature, and in addition has the advantage (as opposed to, e.g., only exponentially increasing SFHs) that it is also applicable to lower redshift galaxies and high-redshift quiescent galaxies, that clearly formed stars at a higher rate in their past than at the epoch of observation.  Maraston et al. (2010) showed that adopting exponentially increasing SFHs for star-forming galaxies at $z \sim 2$ leads to negligible changes in the estimated stellar masses, unless also a stringent constraint on formation redshift is imposed, in which case the estimated stellar mass of these galaxies can increase by a few 0.1 dex.  Solar metallicities were assumed throughout, and dust attenuation was modeled using the Calzetti et al. (2000) reddening law, with visual extinctions in the range $0 < A_V < 4$.  For a detailed description of the dependence of estimated stellar masses on the assumed metallicity and attenuation law, we refer the reader to Wuyts et al. (2007).  We revisit the assumption of solar metallicity in Section\ \ref{model_IRUV.sec} in the context of a model for IR/UV ratios that involves a metallicity scaling, and conclude that it has a negligible impact on our results.  As for the SDSS galaxies, we assumed a universal Chabrier (2003) IMF for all galaxies in the lookback surveys.

\subsubsection{Star Formation Rates}

Exploiting deep PACS photometry in the GOODS-South field, Wuyts et al. (2011) established a continuity across SFR indicators, ranging from UV + FIR, over UV + MIR, to SED modeled SFRs.  We adopt this cross-calibrated 'ladder of SFR indicators' to obtain an estimate of the SFR for each galaxy in our sample individually.  Briefly, if the galaxy is detected in an IR band, we compute the total SFR by summing the unobscured and re-emitted emission from young stars, following Kennicutt (1998):

\begin {equation}
SFR_{UV + IR}\ [M_{\sun} yr^{-1}] = 1.09 \times 10^{-10}\ (L_{IR} + 3.3 L_{2800})/L_{\sun}
\label{SFRuvir.eq}
\end {equation}

where $L_{2800} \equiv \nu L_{\nu}(2800\AA)$ was computed with EAZY from the best-fitting SED.  Here, the total IR luminosity $L_{IR} \equiv L(8-1000\mu m)$ is derived monochromatically from the longest wavelength IR band that is significantly ($> 3\sigma$) detected: either PACS 160, 100, or 70 $\mu$m, or MIPS 24 $\mu$m.  We use the mid- to far-infrared SED template by Wuyts et al. (2008) for the conversion to $L_{IR}$.  This conversion leads to a consistency between 24 $\mu$m and PACS-derived $L_{IR}$, unlike locally calibrated template sets that have been proven to overestimate the $L_{IR}$ of $z \gtrsim 2$ galaxies based on their rest-frame 8 $\mu$m emission (Nordon et al. 2010; Elbaz et al. 2010; Wuyts et al. 2011).

For galaxies that lack an IR detection, we adopt the SFR associated to the best-fit stellar population synthesis model from Section\ \ref{SEDmodeling.sec}.  For the low- to intermediate SFR regime, where $SFR_{UV+IR}$ is only attainable in the deepest of our four fields (GOODS-South), Wuyts et al. (2011) demonstrated that SED modeled SFRs are consistent with those obtained from Equation\ \ref{SFRuvir.eq}.  As for the stellar masses, the values of $SFR_{SED}$, obtained without imposing explicit constraints on the formation redshift, would not change significantly when assuming exponentially increasing rather than decreasing SFHs (Maraston et al. 2010).

\subsubsection{Structural Parameters}
\label{structural_params.sec}

In this paper, we limit our analysis of galaxy structure to a parametric approach, focussing on two basic properties of the surface brightness distribution: its extent and cuspiness.  The former is expressed as the circularized effective radius $r_e \equiv r_{e, major} \times \sqrt{b/a}$ corresponding to the best-fit Sersic profile.  The latter is quantified by the Sersic index $n$, with $n$ restricted to the range $0.2 < n < 8$.  We adopt the structural measurements performed on the longest wavelength high-resolution imaging available for each field (WFC3 for GOODS-S and UDS, ACS for GOODS-N and COSMOS, see Section\ \ref{fields_data.sec}).  The GALAPAGOS package (Barden et al. 2009) or equivalent procedures were used to pre-determine the sky background and the neighboring sources that need to be simultaneously fit or masked, and to automate fitting Sersic profiles to the 2D surface brightness distributions with GALFIT (C. Y. Peng et al. 2010).  GALFIT takes into account convolution by the point spread function (PSF).

\begin {figure*}[t]
\centering
\plotone{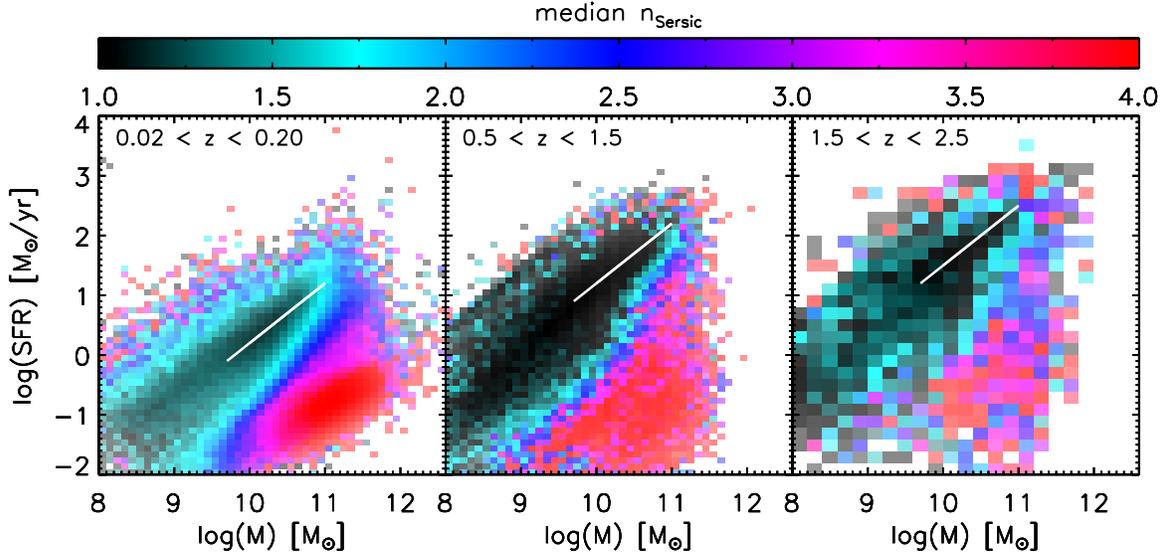} 
\caption{
Surface brightness profile shape in the SFR-Mass diagram.  A 'structural main sequence' is clearly present at all observed epochs, and well approximated by a constant slope of 1 and a zeropoint that increases with lookback time ({\it white line}).  While SFGs on the MS are well characterized by exponential disks, quiescent galaxies at all epochs are better described by De Vaucouleurs profiles.  Those galaxies that occupy the tip and upper envelope of the MS also have cuspier light profiles, intermediate between MS galaxies and red \& dead systems.
\label{n.fig}}
\vspace{-0.12in}
\end {figure*}

Since the longest wavelength at which high-resolution imaging is presently available varies between our four deep fields, our study lacks the potential to consistently probe the rest-frame optical regime, and fully control morphological k-corrections, if present.  However, since we are interested not only in galaxies on the main sequence, but also in those rare systems that form the high-SFR tail, we prefer to sacrifice this aspect of our analysis, rather than area.  As a sanity check, we compared sizes and Sersic indices measured in the $z_{850}$ and $H_{160}$ imaging of GOODS-S (see Appendix B.2).  We find the median ACS - WFC3 deviations to be small (-0.01 in $\Delta \log r_e$ and -0.10 in $n$ at $1.5 < z < 2.5$) relative to the trends that will be discussed in this paper, suggesting that any biases from morphological k-corrections have only a limited impact on our conclusions.  Bond et al. (2011) arrive at a similar conclusion comparing rest-frame optical and rest-frame UV size measurements of $z \sim 2$ galaxies.  Moreover, we also verified that the results obtained for each of the four fields individually are consistent with those presented for the combined data set in this paper, albeit with more noise due to the smaller number statistics (see Appendix B.1).

While beyond the scope of this paper, the wavelength dependence of structural properties can potentially reveal interesting clues on the physical processes shaping galaxies in the young universe (see, e.g., Guo et al. 2011; Szomoru et al. 2011; and Cassata et al. 2011, who find evidence for mild morphological k-corrections such that the centers of galaxies in their samples tend to be somewhat redder).  For an in depth discussion of resolved stellar populations inferred from spatial variations in color, we defer the reader to Wuyts et al. (in prep).

\subsection{Sample Selection}
\label{sample.sec}

Our four deep fields are not uniform in depth.  Consequently, they have different completeness limits in the SFR-Mass diagram.  Since we are mainly interested in individual galaxy characteristics rather than abundances (number or mass densities) as function of position along or across the MS, we refrain from applying any incompleteness corrections to the observed populations.  Under the premise that, at any given redshift, galaxies form a two-parameter family described by their mass and SFR, this approach works well.  I.e., a median galaxy property can be computed reliably, and unbiased by any completeness issues, based on the objects observed in a given bin of SFR-Mass space.

Our final sample comprises 639924 galaxies at $0.02 < z < 0.2$, 132328 galaxies at $0.5 < z < 1.5$, and 35649 galaxies at $1.5 < z < 2.5$.  The relative breakdown in galaxies of different masses is determined by the depth of the observations, and the stellar mass function at the respective redshifts.  Above $M > 10^{10}\ M_{\sun}$ our sample counts 532131, 31127, and 8895 galaxies at $z \sim 0.1$, $z \sim 1$, and $z \sim 2$ respectively.  Above $M > 10^{11}\ M_{\sun}$, the numbers drop to 147922, 2767, and 1059 galaxies at $z \sim 0.1$, $z \sim 1$, and $z \sim 2$ respectively.  An overview of the sample size per field is provided in Table\ \ref{fields.tab}.

\section {Results on Galaxy Structure}
\label{structure.sec}

\subsection {Profile Shape}
\label{profile.sec}

\begin {figure*}[t]
\centering
\plotone{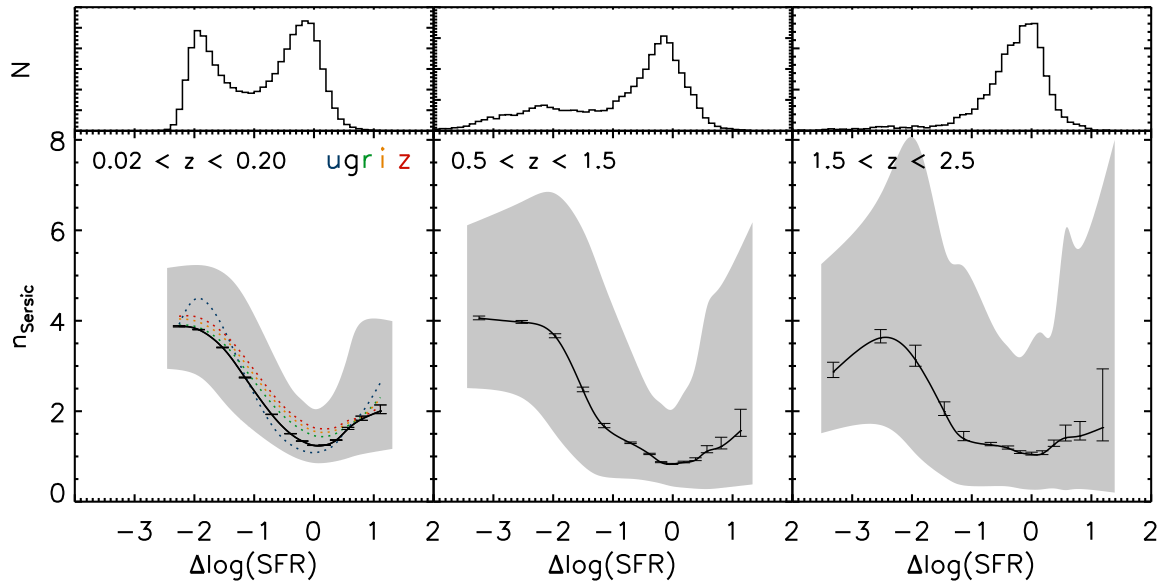} 
\caption{
{\it Top panels:} SFR histograms relative to the 'structural main sequence' for galaxies in the mass range $5 \times 10^{9}\ M_{\sun} < M < 10^{11}\ M_{\sun}$.  {\it Bottom panels:} Sersic index as function of deviation from the MS, measured along the SFR axis, for the same $10^{9.7} - 10^{11}\ M_{\sun}$ mass slice.  The black curve and error bars indicate the median in bins of $\Delta \log (SFR)$, and respective errors in the median.  The grey polygon marks the central 68th percentile of the distribution.  Colored dotted curves in the left-hand panel illustrate how the $z \sim 0.1$ relation changes when adopting structural measurements performed at shorter/longer wavelengths than the SDSS $g$ band.  Relative to the MS, a correlation between profile shape and star-forming activity is present and looks similar at all epochs.  The relation is not monotonic, but rather shows a reversal at the high-SFR end.
\label{dev.fig}}
\vspace{-0.12in}
\end {figure*}

We start by analyzing the surface brightness profile shape as a function of position in the SFR-Mass diagram in Figure\ \ref{n.fig}.  The three panels show from left to right the $z \sim 0.1$, $z \sim 1$, and $z \sim 2$ bins respectively.  Instead of indicating the relative abundance of galaxies in different regions of the diagram, we use the color-coding to mark the median value of the Sersic index $n$ of all galaxies in each $[SFR, M]$ bin.  For displaying purposes, we restrict the range of the colorbar to $1<n<4$, and assign the same color as $n=1$ and $n=4$ to bins with median $n < 1$ or median $n > 4$ respectively.  The fractions $(f_{n<1};\ f_{n>4})$ of galaxies lying outside these bounds amounts to (0.09; 0.24), (0.41; 0.14), and (0.41; 0.16) at $z \sim 0.1$, $z \sim 1$, and $z \sim 2$ respectively.  The fraction of $[SFR, M]$ bins with median n outside this range is small: (0.02; 0.11) at $z \sim 0.1$, (0.14; 0.15) at $z \sim 1$, and (0.11; 0.11) at $z \sim 2$.  The resulting diagrams present a remarkably smooth variation in the typical galaxy profile shape across the diagram.  Moreover, despite the loss of information on number densities, the so-called main sequence of star formation is immediately apparent, and its presence persists out to the highest observed redshifts.  This 'structural main sequence' consists of galaxies with near-exponential profiles ($n \approx 1$) and shows a similar behavior as the conventional 'number main sequence' as identified on the basis of number densities in the SFR-Mass diagram (e.g., Noeske et al. 2007; Elbaz et al. 2007; Daddi et al. 2007).  Namely, an upward shift of the zeropoint is observed with increasing lookback time.  At each epoch, the MS in Figure\ \ref{n.fig} is well approximated by a slope of unity ({\it white line}).  The SFR at which the median n reaches a minimum in a mass slice around $\log(M) = 10$ roughly coincides with the mode of the $\log(SFR)$ distribution in that mass slice, but depending on the fitting method and sample definition used to weed out quiescent galaxies, a somewhat shallower slope than unity may be measured for the 'number main sequence' at the massive end (see, e.g., Rodighiero et al. 2010).

Below the structural MS, a cloud of galaxies with cuspy, near-De Vaucouleurs ($n \approx 4$) profiles is visible.  This population of massive quiescent galaxies is present at all observed epochs.  Our first and foremost conclusion from Figure\ \ref{n.fig} is therefore that a correlation between the structure and stellar populations of galaxies is already in place since $z \sim 2.5$.  In the local universe, the presence of such a correlation has been described in depth by, e.g., Kauffmann et al. (2003) and Brinchmann et al. (2004).  Scarlata et al. (2007) demonstrated that photometrically selected massive quiescent galaxies are already dynamically relaxed (i.e., have the morphological appearance of early-type galaxies) at $z \sim 1$.  Also out to $z \sim 1$, Maier et al. (2009) exploited the spectroscopic zCOSMOS survey to reveal a correlation between the specific SFR of galaxies and the Sersic index of their surface brightness profiles.  At larger lookback times, our results are in agreement with previous findings, based on smaller samples, reporting a similarly early emergence of the Hubble sequence on the basis of stellar surface mass density, galaxy size and visual appearance (Franx et al. 2008; Toft et al. 2009; Kriek et al. 2009b).  We find that, across cosmic time, the typical Sersic index of galaxies is not optimally described as a function of their absolute SFR, or even their absolute specific SFR, but rather as a function of their position relative to the MS at the epoch of observation.  The correspondence between mass, SFR and structure, as quantified by the Sersic index, is equivalent to the Hubble sequence.  Based on samples of unprecedented size, we see that such a sequence already existed at $z \sim 2$; bulge-dominated morphologies go hand in hand with a more quiescent nature.  First indications of such a correlation between Sersic index and specific SFR out to high redshift were recently reported by Szomoru et al. (2011) based on 27 galaxies at $z \sim 1$ and 16 galaxies at $z \sim 2$.

\begin {figure*}[t]
\centering
\plotone{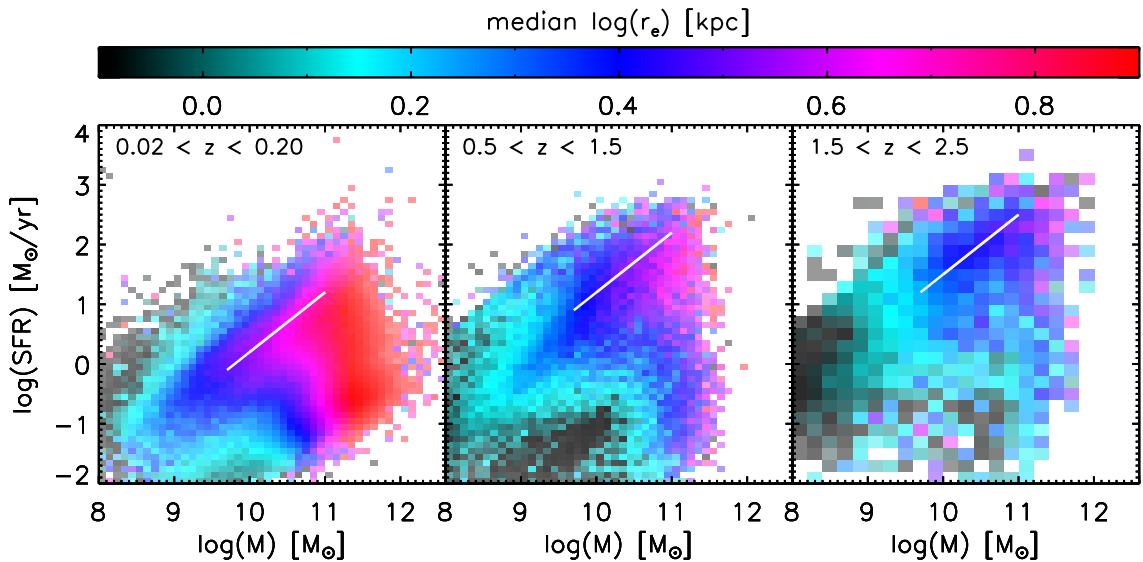} 
\caption{
Galaxy size in the SFR-Mass diagram.  Moving up the main sequence, towards the high-mass end, galaxies grow bigger (i.e., a size-mass relation is observed at all epochs).  With respect to both quiescent galaxies, and also the high-SFR tail, galaxies on the MS have the largest size.  Finally, in any given $[\log (M); \log (SFR)]$ bin, galaxies grow with time.  Quantitatively, this growth is strongest for quiescent (low $SFR/M$) galaxies.
\label{re.fig}}
\vspace{-0.12in}
\end {figure*}

In detail, an additional interesting aspect is revealed by the $n$ distribution in SFR-Mass space.  The surface brightness profile shape does not vary monotonically across the main sequence, but instead shows a reversal in the high-SFR tail of the distribution, such that the upper envelope of the MS is composed of galaxies whose typical cuspiness is intermediate between those of normal MS galaxies and the quiescent population below the MS.  This is illustrated further in Figure\ \ref{dev.fig}, where we plot a cut through of the $5 \times 10^{9}\ M_{\sun} < M < 10^{11}\ M_{\sun}$ mass slice.   The top panels show the histogram of deviations from the MS measured along the SFR axis: $\Delta \log(SFR)$.  The top left panel illustrates the well-known bimodality between star-forming and passive galaxies in the local universe (e.g., Kauffmann et al. 2003).  A similar bimodality, albeit with a reduced amplitude of the quiescent peak, is visible at $z \sim 1$ (and becomes more pronounced if we were to limit ourselves to higher masses only).  While both populations are also found in our highest redshift bin, the quiescent fraction is further reduced.  Incompleteness in our shallowest field, as well as uncertainties in photometric redshifts and estimated SFRs inhibit the identification of a true bimodality in the $z \sim 2$ histogram.

The bottom panels of Figure\ \ref{dev.fig} show the median dependence of $n$ on $\Delta \log(SFR)$ ({\it black curve}), as well as errors on the determination of the median ({\it black error bars}).  With grey polygons, we mark the central 68th percentile.  Clearly, a variety of profile shapes is measured, even at a given $\Delta \log (SFR)$, and particularly in the $z \sim 1$ and $z \sim 2$ bins.  This stems partly from measurement uncertainties, that dominate in the low surface brightness regime, but in addition likely means that in detail galaxies are more complex than a simple two-parameter ($[SFR, M]$) family.  Interestingly, the scatter seems to be minimized for galaxies that reside on the MS.  F\"{o}rster Schreiber et al. (2011) present a detailed analysis of large and clumpy MS galaxies at $z \sim 2$, that exhibit shallow surface brightness profiles ($n < 1$).  Such a population is also present in our sample (see Figure\ \ref{dev.fig} and the above quoted $f_{n<1}$).  Focussing on the overall trend, we find the results at $z \sim 1$ and $z \sim 2$ to be very similar to those established at $z \sim 0.1$.  Namely, galaxies on the MS resemble exponential disks, high-SFR outliers have a centrally enhanced surface brightness profile, and once galaxies end up below the MS, they quickly reach a plateau of $n \approx 4$.  We discuss the physical implications of this empirical relation in Section\ \ref{quenching.sec}.

We verified that, when excluding X-ray sources from our analysis, the same median dependence of the surface brightness profile shape on the location of galaxies in the SFR-Mass diagram remains present.  This is the case even for the fields with the deepest X-ray data.  The same robustness also applies to the other galaxy properties discussed later in this paper.  Our test indicates that the observed trends (particularly the enhanced Sersic indices of the high-SFR outliers at a given mass) is not merely due to a nuclear, non-stellar point source.  When focussing on the X-ray detected sources alone, we find that they occupy the mass regime above $\log M \gtrsim 10$, and span a wide range in SFRs, from the quiescent class to on and above the MS.  Within a given bin, their surface brightness profiles differ in the sense that they have higher Sersic indices and smaller sizes than normal galaxies of the same $[SFR, M]$.  Point source contamination would drive the structural parameters in the observed direction, but extensive simulations (see, e.g., Simmons \& Urry 2008) are required to constrain whether in addition there is any evidence for a different stellar structure of the hosts.

Before moving to the next structural property, we have a closer look at the $z \sim 0.1$ bin in Figure\ \ref{dev.fig}, where we use structural measurements in five passbands to investigate how the observed trend depends on wavelength.  The dotted colored curves in the bottom left panel of Figure\ \ref{dev.fig} illustrate the median relation derived from Sersic fits in the SDSS $u$, $r$, $i$, and $z$ bands.  Together with the default $g$-band profile fits ({\it solid black curve}), a trend emerges in which the overall relation between $n$ and $\Delta \log (SFR)$ is remarkably robust against morphological k-corrections.   The largest difference is observed for galaxies that lie on the MS ($\Delta \log (SFR) = 0$), such that a cuspier profile is inferred from redder waveband imaging.  This comes as no surprise, as it is well known that many normal SFGs in the nearby universe are composed of a (red, $n \approx 4$) bulge and (blue, $n \approx 1$) disk component\footnote{The relative fraction of such bulged disks is observed to decrease with redshift, at least out to $z \sim 1$ (Sargent et al. 2007).}.  At low and high values of $\Delta \log (SFR)$, the Sersic index levels of at similar values ($n = 4$ and $n = 2$ respectively), irrespective of wavelength.  This suggests that the light concentration in the centers of these galaxies corresponds to a mass concentration, and is not due (only) to the large luminosity of a young stellar population.

\subsection {Size}
\label{size.sec}

\begin {figure*}[t]
\centering
\plotone{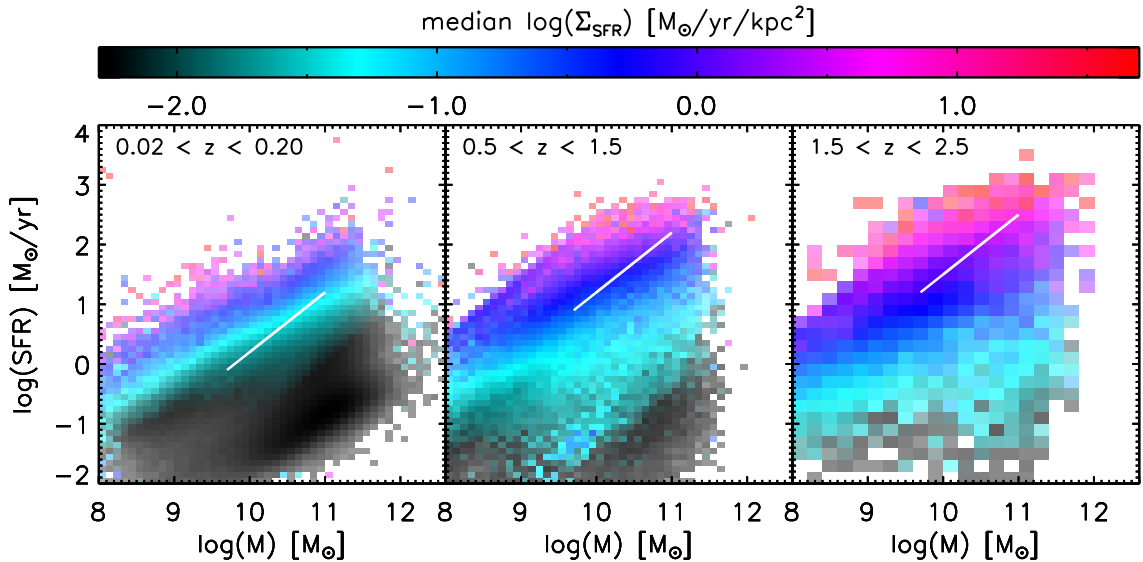} 
\caption{
Surface density of star formation ($\Sigma_{SFR}$) in the SFR-Mass diagram.  Lines of equal $\Sigma_{SFR}$ run approximately parallel to the MS, as do lines of constant $SFR/M$. 
\label{surf.fig}}
\vspace{-0.12in}
\end {figure*}

Having established the existence of a 'Hubble sequence' out to $z \sim 2.5$, we now turn to the zeroth order structural measurement of size, and address how it varies along and across the MS at each of our observed epochs.  Figure\ \ref{re.fig} is similar in nature to Figure\ \ref{n.fig}, but the color-coding is now used to mark differences in the spatial extent of the surface brightness distribution ($r_e$).  As in Figure\ \ref{n.fig}, a qualitatively similar behavior is observed at all redshifts.  Overall, more massive galaxies tend to be larger.  However, the mass-size relation is not fundamental in the sense that at a given mass substantial variations in size occur that are not random, but instead correlate with the offset from the MS (see also the Appendix, Figure\ \ref{dev_app.fig}).  This translates to different mass-size relations followed by passive and star-forming galaxies, as reported previously by Shen et al. (2003) based on SDSS, but also by Williams et al. (2010) and Szomoru et al. (2011) who extended the analysis for massive galaxies out to $z = 2$, based on ground-based $K$-band imaging in UDS and WFC3 $H_{160}$-band imaging in the Hubble Ultra Deep Field respectively.

Consistent with previous results by, e.g., Franx et al. (2008) and Toft et al. (2009), high-redshift quiescent galaxies are more compact than the bulk of SFGs at the same epoch.  Also at $z \sim 0.1$, galaxies below the MS (at least at $M < 10^{11}\ M_{\sun}$) tend to be smaller than those on the MS.  The high-SFR tail of the $z \sim 0.1$ distribution differs from normal MS galaxies in the sense that they tend to have higher surface mass densities (i.e., are more compact for a given mass).  As we march up in redshift, our measurements are consistent with a similar behavior across all lookback times considered.

Comparing the panels for the three redshift bins, it is immediately apparent that galaxies grow over time.  We quantified the growth by fitting a $r_e(z) / r_e(z=0) = (1+z)^\alpha$ relation to the galaxies in each $[SFR, M]$ bin.  It is important to note that this reflects the size evolution of the galaxy population as a whole, and not per se that of any given galaxy individually, as galaxies will move from one $[SFR, M]$ bin to another as they build up stellar mass, and new systems will be added over time.  We find $\alpha$ to be varying across the SFR-Mass diagram, from $\alpha = -1.2_{-0.4}^{+0.9}$ below $\log (SFR/M) < -11$ to $\alpha = -0.4_{-0.7}^{+0.7}$ above $\log (SFR/M) > -11$.   Here, the range indicates the normalized median deviation over the different $[SFR, M]$ bins.  Splitting the latter class into galaxies above and below the MS at $z \sim 0.1$ ({\it white line in the left-hand panel of Figure\ \ref{re.fig}}), we obtain $\alpha = -0.2_{-0.6}^{+0.5}$ and $\alpha = -0.9_{-0.5}^{+0.6}$ respectively.  In other words, the rate of size evolution correlates with the specific star formation rate, reaching the lowest values in the (especially massive) quiescent population.

\section{Results on Star Formation Mode}
\label{mode.sec}

Having established the variation of galaxy structure along and across the MS, we now do the same for two parameters that characterize the mode of star formation: its surface density (Section\ \ref{surfdens.sec}), and its breakdown into unobscured and dust-enshrouded star formation (Section\ \ref{obscuration.sec}).  A discussion of the physical implications of the observed relationships follows in Section\ \ref{model_IRUV.sec}.

\subsection{Surface Density of Star Formation}
\label{surfdens.sec}

Figure\ \ref{surf.fig} presents the dependence of the SFR surface density

\begin {equation}
\Sigma_{SFR} \equiv SFR / 2\pi r_e^2
\label{SFRsurfdens.eq}
\end {equation}

on the position of galaxies in the SFR-Mass diagram.  By construction, this Figure has the same information content as Figure\ \ref{re.fig}, but its interpretation in terms of mode of star formation is more straightforward.  Galaxies that form stars more actively are not just upscaled versions of lower SFR systems.  Their ISM conditions differ in the sense that also normalized by area they form more stars per unit time.  Lines of constant $\Sigma_{SFR}$ run diagonally, implying a tighter correlation with specific SFR than with SFR or stellar mass separately.  Schiminovich et al. (2007) discussed this trend extensively for galaxies in the nearby universe.  At higher redshifts, this is consistent with recent findings based on rest-frame UV size measurements by Elbaz et al. (2011).  We confirm that the same trend is seen when rest-frame optical size measurements are used instead.

\begin {figure}[htbp]
\centering
\plotone{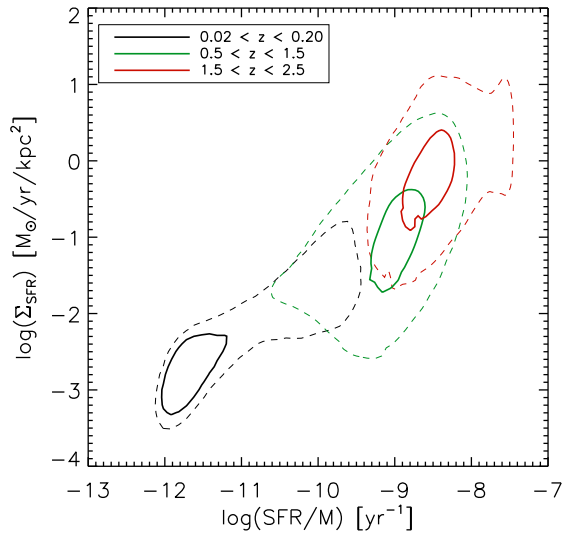} 
\caption{Surface density of star formation as function of specific star formation rate.  Solid and dashed contours comprise 25\% and 75\% of the galaxies respectively.  At all observed epochs, galaxies line up along the same fundamental relation, but towards higher redshifts the bulk of galaxies shifts to the higher $\Sigma_{SFR}$ (or $SFR/M$) end of the relation. 
\label{surfSSFR.fig}}
\vspace{-0.12in}
\end {figure}

We contrast $\Sigma_{SFR}$ to $SFR/M$ in Figure\ \ref{surfSSFR.fig}, where we plot the 25th and 75th percentile contours ({\it solid and dashed lines}) for each of the considered redshift bins.  Over 10 Gyr of lookback time, the bulk of galaxies follow a similar, linear relation between specific SFR and surface density of star formation.  As we probe higher redshifts, the upper end of this relation becomes more densily populated.  This trend can naturally be explained by increased gas mass fractions in the young universe (e.g., Tacconi et al. 2010).

\begin {figure*}[t]
\centering
\plotone{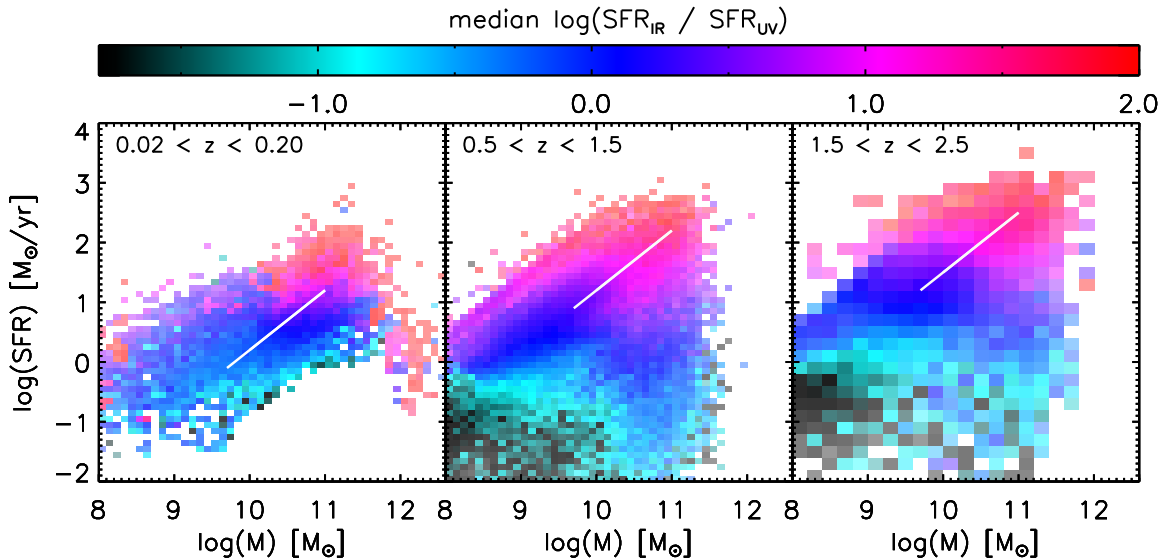} 
\caption{
Ratio of re-emitted over unobscured star formation in the SFR-Mass diagram.  The highest $SFR_{IR}/SFR_{UV}$ ratios are found at the upper rim, and at the high-mass end of the MS.
\label{IRUVobs.fig}}
\end {figure*}

In equation\ \ref{SFRsurfdens.eq}, we adopted the same size measurement as discussed in Section\ \ref{structural_params.sec} and\ \ref{size.sec} to represent the radius within which half the star formation takes place.  I.e., we adopted the best-fit $r_e$ of a single Sersic profile fit.  At $z \sim 0.1$, we used the $g$ band by default.  For the higher redshift samples, we used the longest wavelength high-resolution image available ($H_{160}$ in the case of UDS and GOODS-S, $z_{850}$ and $I_{814}$ for GOODS-N and COSMOS respectively).  In principle, radial variations in SFH and/or dust attenuation could bias the measurement of a half-SFR radius.  E.g., in the case of a superposition of a star-forming disk and a red and dead bulge, the total half-light radius would contain less than half the star formation (particularly when measured in the red).  In contrast, in the case of a galaxy where old and young stars trace the same distribution, but the dust column is centrally enhanced, the observed $r_e$ would contain more than half of the star formation (particularly when measured in the blue).  We take two simple approaches to investigate the impact of these potential biases on our results.  First, we repeated our analysis at $z \sim 0.1$ adopting each of the $ugriz$ SDSS bands to determine $r_e$ and based thereupon $\Sigma_{SFR}$.  We find that this induces a minor change only, in the sense that, for galaxies below the main sequence, $\Sigma_{SFR}$ increases by a few 0.1 dex from the bluest to the reddest band (see the left-hand panel of Figure\ \ref{dev_app.fig} in the Appendix).  Overall, the pattern of $\Sigma_{SFR}$ in SFR-Mass space does not depend appreciably on the adopted photometric band at $z \sim 0.1$.  Second, we stress that at $z \sim 1$ as well as at $z \sim 2$, the dependence of $\Sigma_{SFR}$ on location in the SFR-Mass diagram remains the same (although with somewhat enhanced noise due to the smaller number statistics) when considering the fields with rest-frame optical and rest-frame UV imaging separately (see Appendix B.1).

\subsection{Obscuration of Star Formation}
\label{obscuration.sec}

Finally, we consider the ratio of re-emitted to unobscured star formation.  What we denote as '$SFR_{IR}/SFR_{UV}$' is for IR-detected galaxies computed as the ratio of the IR and UV contribution to the total $SFR_{UV+IR}$ (see Equation\ \ref{SFRuvir.eq}): $L_{IR} / 3.3 L_{2800}$.  For those galaxies in the deep lookback surveys that lack an IR detection, and for which we adopted the best-fit SFR from stellar population modeling, we derive $SFR_{IR}/SFR_{UV}$ as $10^{0.4 A_{2800}} - 1$.  Here, $A_{2800}$ is the extinction at 2800\AA\ as inferred from the SED modeling.  This approach has the advantage of always producing positive values, unlike adopting $(SFR_{SED} - SFR_{UV} )/ SFR_{UV}$ as definition, which can reach negative values when $SFR_{UV} > SFR_{SED}$.  The latter situation can occur for quiescent galaxies with low ratios of present to past-averaged SFR, in which case the rest-frame 2800\AA\ emission contains a substantial contribution from older stars.

For the $z \sim 0.1$ sample, total SFRs (based on dust- and aperture-corrected $H\alpha$ measurements, see Section\ \ref{SDSS_GALEX.sec}) are available from the MPA-JHU release, and we computed $SFR_{UV}$ as for the higher redshift galaxies, where we interpolated $L_{2800}$ from the GALEX NUV and SDSS $u$-band photometry.  The derived $SFR_{IR}/SFR_{UV} = (SFR_{tot} - SFR_{UV} )/ SFR_{UV}$ at $z \sim 0.1$ are well defined for the star-forming population, but ill-constrained for the passive population for the above reasons.  We therefore choose to exclude the passive $z \sim 0.1$ galaxies from our analysis.

The resulting SFR-Mass diagrams, color-coded by $SFR_{IR}/SFR_{UV}$,  are presented in Figure\ \ref{IRUVobs.fig}.  Again, the qualitative appearance of the diagrams looks similar over the wide range of redshifts considered.  At a given mass, a proportionally larger fraction of the star formation activity is revealed by reprocessed IR emission as we consider more actively star-forming systems.  A positive correlation between bolometric luminosity and obscuration was presented previously for galaxies in the nearby universe (Heckman et al. 1998) and at $z \sim 2$ (Reddy et al. 2010), but these authors did not trace the trend with SFR while specifically controlling for stellar mass.   Likewise, an increase in $SFR_{IR}/SFR_{UV}$ is observed as we move along the MS from the low- to high-mass end.  Such a trend was not seen as clearly in $\Sigma_{SFR}$ (Figure\ \ref{surf.fig}), a finding that we will interpret as due to metallicity variations across the diagram in Section\ \ref{model_IRUV.sec}.  Finally, objects forming the upper rim of the galaxy distribution, also at lower masses, show enhanced IR emission.  The latter trend is most prominent in the $z \sim 1$ bin, but a hint of the same feature is seen at both lower and higher redshift as well.

\section {Discussion}
\label{discussion.sec}

\subsection{The Relation between Peak Starbursts and Quiescent Galaxies}
\label{quenching.sec}

The observation that a precursor of the Hubble sequence, with disk-like SFGs and cuspier quiescent systems, was already in place 10 Gyr ago (see Section\ \ref{profile.sec}) has important implications.  To name one, a mechanism for galaxies to transition from the star-forming to quiescent phase must be universally present across cosmic time, at least out to $z \sim 2.5$.  It is during this epoch around $z \sim 2$ that the cosmic SFR density is observed to reach its peak (e.g., Hopkins \& Beacom 2006), and that large accretion rates onto galaxy halos are expected from cosmological hydrodynamical simulations (Keres et al. 2005, 2009; Dekel \& Birnboim 2006; Dekel et al. 2009).  We can therefore exclude gas exhaustion as sole driver of the transition from the active to passive phase.  Instead, a quenching mechanism must be at play.  Moreover, the observed relation between structure and stellar populations implies a causal connection between the quenching and the morphological transition.

As early as by Toomre \& Toomre (1972), mergers have been invoked as a potential mechanism to transition from late to early Hubble types.  In such a scenario, much of the gas content of the progenitors is consumed by star formation in the nucleus, building up a central cusp.  Violent relaxation distributes the stars that were already formed prior to final coalescence in a near-De Vaucouleurs ($n=4$) distribution.  Feedback from short-lived massive stars that explode as supernovae, as well as energy that is released from accreting material onto the SMBH(s) and that couples to the surrounding ISM drives out the remaining gas and heats up the galaxy halo preventing new infall (e.g., Hopkins et al. 2006, and references therein).

In this light, the reversal of the decreasing trend of $n$ with $\Delta \log(SFR)$ (see Figure\ \ref{dev.fig}) may signal that we are observing the galaxies at the high-SFR tail of the distribution during the peak of their SFH, in the process of building up a central concentration of stellar mass, much like that typically observed in the quiescent population.  This would also explain their increased SFR surface densities and small sizes relative to normal SFGs on the MS.  While galaxies above and below the MS by definition form the opposite extremes in terms of SFR, they may thus share a more similar nature in terms of structural shape.  According to the merger evolutionary sequence, this would place them as nuclear starbursts in between MS galaxies and quiescent remnants.

A crude duty cycle argument reveals that, if all quiescent galaxies at $0.5 < z < 1.5$ with $10^{10}\ M_{\sun} < M < 10^{11}\ M_{\sun}$ and $\Delta \log (SFR) < -1$ underwent a starburst of $\Delta \log (SFR) > 0.5$ prior to quenching, the timescale of that burst must have been of order 110 Myr to account for the observed numbers of such bursts:

\begin {equation}
\Delta t_{burst} = \frac {n_{burst}}{n_{new\ quiescent}} \Delta t_{0.5<z<1.5}
\label {duty.eq}
\end {equation}

Here, $n_{burst}$ represents the number density of objects with $\Delta \log SFR > 0.5$, and $n_{new\ quiescent}$ equals the number of newly quenched ($\Delta \log (SFR) < -1$) systems since $z = 1.5$ (which is most of them, given the relatively low number density at $z > 1.5$).  The time interval probed in the $z \sim 1$ bin spans $\Delta t_{0.5<z<1.5} = 4.2$ Gyr.  For the $z \sim 2$ bin, similar arguments lead to a value of $\sim 70$ Myr.  To determine these timescales, we restricted ourselves to the deepest fields, GOODS-South and UDS, to avoid any incompleteness effects.  Our WFC3 $H_{160}$-selected samples in these fields do not suffer from incompleteness above $10^{10}\ M_{\sun}$, even for maximally old, high $M/L$ systems in our highest redshift bin (see Appendix B.1).  Increasing the number statistics by pushing down to $5 \times 10^9\ M_{\sun}$, where we are still highly complete, we obtain timescales of 170 Myr and 120 Myr at $z \sim 1$ and $z \sim 2$ respectively.  The derived timescales should be considered as crude upper limits.  First, uncertainties in SFR estimates may lead to a contamination in the high-SFR tail by galaxies with intrinsically lower SFR.  Second, in the case that not all excursions to high SFRs are followed by quenching, the inferred timescale of bursts associated with a quenching event will be shortened further.  Overall, the timescales obtained are not inconsistent with those expected from merger-induced star formation on the basis of hydrodynamical simulations (Mihos \& Hernquist 1996; Cox et al. 2008), although there may be room for other evolutionary paths to reach the quiescent phase as well.

Our observations of the structural MS and the quiescent population located below in Figure\ \ref{n.fig} reveal two aspects that any galaxy formation and quenching model should account for.  First, we find spheroid-like ($n \sim 4$) quiescent galaxies to be present over a wide range of stellar masses (1.5 - 2 dex) at all redshifts, down to at least $10^{10}\ M_{\sun}$.  Second, galaxies with a large spread in star formation activities co-exist at the same stellar mass over a mass range of an order of magnitude (roughly $10 < \log (M) < 11$).  If all galaxies at a given stellar mass are central galaxies and reside in an equally massive dark matter halo, this would imply that quenching is not a simple step function of halo mass.  Carrying out a detailed abundance matching analysis between galaxies at $0 < z < 1$ and their halo hosts, Conroy \& Wechsler (2009) also do not find evidence for a sharp characteristic halo mass at which star formation truncates.  If on the other hand halo quenching is responsible for the transition from active to passive systems, this would imply that quiescent galaxies live in higher mass halos (i.e., have a smaller baryon fraction) than SFGs of similar stellar mass.

Finally, a third quenching mechanism that can contribute to the build-up of the quiescent population is environmental (satellite) quenching.  Studying the role of environment and its impact on the morphological properties of (part of) the quenched population, is beyond the scope of this paper.  We do however note that Y. J. Peng et al. (2010) argue this physical process only becomes relevant at late times ($z \lesssim 0.5$) and low masses ($\log M < 10.5$).  Based on an empirical model that successfully reproduces the stellar mass function of star-forming and passive galaxies in SDSS, Y. J. Peng et al. (2010) attribute the quenching at larger lookback times to a combination of a mass-dependent component (dominating at the high-mass end), and a mass-independent merger quenching component.  Together, these processes are claimed to be responsible for the double Schechter mass function of passive galaxies.  It is important to point out that their 'mass quenching' is probabilistic in nature, rather than a step function, with a rate proportional to the SFR, and an underlying physical origin that may well be associated with AGN or star formation feedback rather than a halo-scale phenomenon.

\subsection{A Physical Model for IR/UV Ratios}
\label{model_IRUV.sec}

In Sections\ \ref{surfdens.sec} and\ \ref{obscuration.sec}, we presented observational results on how the surface density and obscuration of star formation depends on the position of galaxies in the SFR-Mass diagram.  Here, we investigate whether those two observations can be understood in the context of one self-consistent physical model that relates the amount of reprocessed emission to the column of obscuring material, and indirectly to $\Sigma_{SFR}$.  The aim of such a model is to explain qualitatively the difference in shape of iso-$\Sigma_{SFR}$ and iso-$SFR_{IR}/SFR_{UV}$ regions in the SFR-Mass plane (Figure\ \ref{surf.fig} versus Figure\ \ref{IRUVobs.fig}), and to reproduce quantitatively the observed $SFR_{IR}/SFR_{UV}$ ratio starting from $\Sigma_{SFR}$, with the help of a few fundamental, and where possible empirically calibrated, laws.

\begin {figure*}[t]
\centering
\plotone{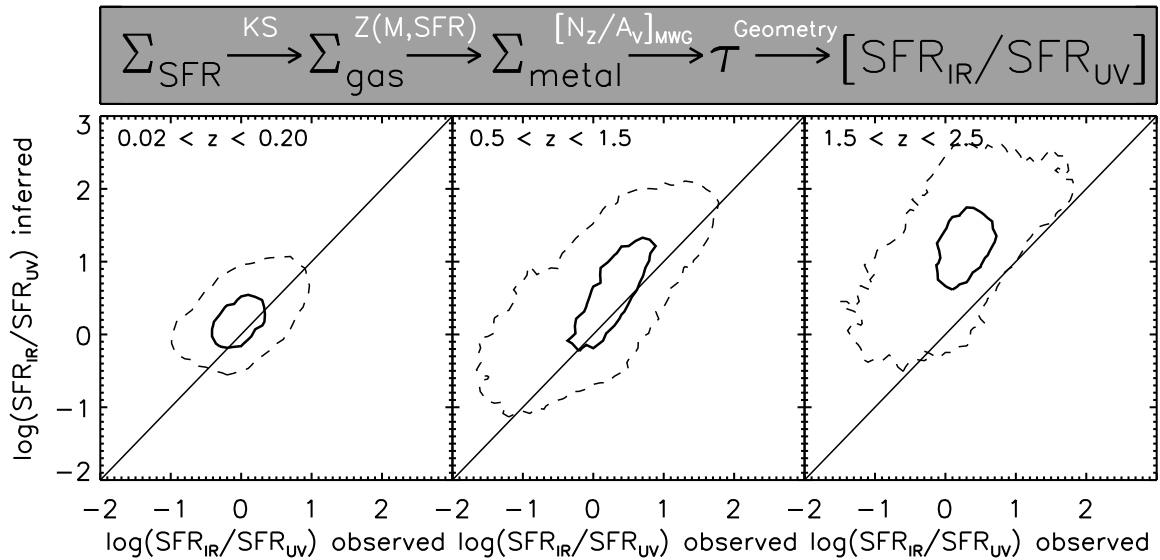} 
\caption{
A comparison between the observed $SFR_{IR}/SFR_{UV}$ ratio and that inferred from the observed $\Sigma_{SFR}$ using a simple physical model that takes into account the SF law, metallicity dependence, and geometric distribution of dust and stars.  The solid and dashed contours mark the 25th and 75th percentiles of the galaxy distribution respectively.  While at $z \sim 0.1$ the model agrees with the observations, towards higher redshift a relatively larger amount of UV radiation is observed than expected from the model.  This may infer shortcomings related to the assumed geometry, or alternatively SF law and/or metallicity dependence.
\label{IRUVcomp.fig}}
\end {figure*}

Our model involves three steps: a conversion from SFR surface density to gas surface density, a conversion from gas column to dust column, and a geometry-dependent translation from dust column to IR-over-UV ratio (see top panel of Figure\ \ref{IRUVcomp.fig}).  The first step boils down to an application of the Kennicutt-Schmidt (KS) star formation law (Kennicutt 1998).  Recently, Genzel et al. (2010) made the case for a KS law of the form

\begin {eqnarray}
\log (\Sigma_{SFR} [M_{\sun} yr^{-1} kpc^{-2}]) = & 1.17 \log (\Sigma_{mol\ gas} [M_{\sun} pc^{-2}]) \nonumber \\
  & - 3.48
\label{KS.eq}
\end {eqnarray}

on which both nearby and high-redshift main sequence SFGs lie.  Major mergers follow a relation of similar slope according to these authors, but may be offset in normalization, so that a higher level of star formation is obtained for a given reservoir of molecular gas.  For simplicity, we assume Equation\ \ref{KS.eq} holds for all galaxies in our sample, at low and high redshift.  We apply Equation\ \ref{KS.eq} to our observed $\Sigma_{SFR}$, account for a 36\% mass correction for Helium to obtain the $H_2$ surface mass density, and add a fiducial plateau of $HI$ surface mass density $\Sigma_{HI} = 4.5\ M_{\sun} pc^{-2}$ (Krumholz, McKee \& Tumlinson 2009) to obtain the total hydrogen column.  In most cases, the contribution of HI to the total gas column is negligible.

The second step of our model accounts for the fact that not all galaxies have the same metallicity.  This is relevant, because the limited number of cases in which metallicities as well as dust-to-gas ratios have directly been observed (Milky Way, Small and Large Magellanic Cloud), illustrate that it is not the dust-to-gas ratio, but rather the dust-to-metal ratio that remains relatively constant between galaxies.  We adopt a gas-to-dust ratio equal to that of the Milky Way $(N_H / A_V)_{MWG} = 1.87 \times 10^{21}\ cm^{-2}\ mag^{-1}$ (Bohlin et al. 1978), with a linear scaling accounting for gas-phase metallicities deviating from solar: 

\begin {equation}
N_H / A_V = (Z / Z_{\sun})^{-1} (N_H / A_V)_{MWG}.
\label{NH_Av.eq}
\end {equation}

Even in the era of multi-object near-infrared spectrographs, spectral metallicity tracers of high-redshift galaxies, such as $[NII]/H\alpha$ are notoriously difficult to obtain for large samples of galaxies, let alone for the sample size considered here.  We therefore adopt an empirically calibrated functional form to compute the metallicity on the basis of known stellar population properties.  Rather than assuming an evolution of the mass-metallicity relation with redshift $Z(M, z)$ (e.g., Erb et al. 2006; Maiolino et al. 2008; Mannucci et al. 2009; Buschkamp et al. in prep), we parametrize the metallicity as a non-evolving function of position in the SFR-Mass diagram: $Z(SFR, M)$.  Using the SDSS spectral database, Mannucci et al. (2010) established such a fundamental surface in SFR-Mass-Metallicity space.  Furthermore, these authors find that galaxies out to $z \sim 2.5$ lie on the same surface as do nearby systems.  The observed evolution of the mass-metallicity relation stems in this picture merely from the fact that galaxies with higher SFRs are more abundantly present at high redshift.  That higher SFR systems at a given mass have lower metallicities is interpreted as evidence for recent gas infall that simultaneously fuels the formation of new stars, and dilutes the metal-content of the interstellar medium.

Even in the local universe, measurements of the gas-phase metallicity (and therefore the normalization of the mass-metallicity relationship) vary by factors of $\sim 3$ depending on the tracer and calibration used  (Kewley et al. 2008).  Given this systematic uncertainty, we choose to renormalize the Mannucci et al. (2010) relation so that it reaches a plateau of solar metallicity ($\log (O/H) + 12 = 8.66$, Asplund et al. 2004) at the high-mass end, so that for galaxies with Milky Way properties ($\sim 10^{11}\ M_{\sun},\ 1\ M_{\sun}/yr$), the Milky Way dust-to-gas ratio applies:

\begin{eqnarray}
\log (O/H) + 12 = & 8.49 + 0.47 (\mu_{0.32} - 10)    &  if\ \mu_{0.32} < 10.2 \nonumber \\
                               & 8.66\ \ \  if\ \mu_{0.32} > 10.5
\label{mannucci.eq}
\end{eqnarray}

where $\mu_{0.32} = \log(M) - 0.32 \log (SFR)$.  For $10.2 < \mu_{0.32} < 10.5$, the smooth polynomial form from Mannucci et al. (2010, equation 4) is applied, downscaled by the same factor.

The scaling of the column with $Z(SFR, M)$ accounts at least in part for the difference in shape of the $\Sigma_{SFR}$ and $SFR_{IR}/SFR_{UV}$ diagrams (Figure\ \ref{surf.fig} and\ \ref{IRUVobs.fig} respectively).  While along the MS little evolution in $\Sigma_{SFR}$ is observed, gas-phase metallicities are expected to increase as we march up the MS.  This leads to an increase in the dust column, and consequently of the reprocessed fraction of the star formation towards the high-mass end of the MS, as observed in Figure\ \ref{IRUVobs.fig}.  In other words, our observations do not
probe metallicities directly, but are consistent with the presence of a mass-metallicity and/or mass-SFR-metallicity relation.\footnote{The assumption of solar metallicity in determining the stellar masses (see Section\ \ref{SEDmodeling.sec}) is in principle inconsistent with the gas-phase metallicity scalings applied in our model.  As a sanity check, we verified that all conclusions drawn in this paper remain valid if we were to adopt a stellar metallicity equal to the gas-phase metallicity according to Mannucci et al. (2010): $Z_* = Z(SFR, M(Z_*))$.  The effect of such a scaling is to slightly skew the SFR-Mass diagram, such that generally lower mass galaxies (and at a given mass particularly those with higher SFRs) shift to masses that are lower by 0 to -0.2 dex.  At metallicities of a half solar, the stellar masses decrease by a few percent in the median, while the median decrease amounts to -0.1 dex at a quarter of solar.}

Using the Calzetti et al. (2000) attenuation law in combination with Equation\ \ref{NH_Av.eq}, we now compute the optical depth $\tau_{2800}$ at the rest-frame 2800\AA\ wavelength where we defined $SFR_{UV}$:

\begin{equation}
\tau_{2800} = (1.79 A_V) / 1.086
\end {equation}

The third and final step of our model relates the optical depth to the ratio of reprocessed over unobscured star formation.  For the geometry of a uniform foreground screen, i.e., where all stars are hiding behind the total dust column, the intrinsic UV emission from young stars is reduced by a factor $e^{- \tau}$.  We quickly dismissed such a geometry, as it is physically implausible, and leads to inferred $SFR_{IR}/SFR_{UV}$ that exceed the observed values by several orders of magnitude.  Clearly, UV emission from young stars is observed, and assuming a more realistic mixture of dust and stars in our model can account for this.  We choose to adopt the simplest possible model, in which the emitting sources and the dust are homogeneously mixed.  In such a configuration, the intrinsic emission is attenuated by a factor $(1 - e^{-\tau}) / \tau$ (McLeod et al. 1993; F\"{o}rster Schreiber et al. 2001).  Rewritten to a functional form for $SFR_{IR}/SFR_{UV}$, this becomes:

\begin {equation}
SFR_{IR}/SFR_{UV} = \frac {\tau_{2800} - 1 + e^{-\tau_{2800}}} {1 - e^{-\tau_{2800}}}
\end {equation}

We present the comparison of our model results to the observed values of $SFR_{IR}/SFR_{UV}$ in Figure\ \ref{IRUVcomp.fig}.  Solid and dashed contours enclose 25\% and 75\% of the galaxy distribution respectively.  In the $z \sim 0.1$ bin, passive galaxies, for which our $SFR_{IR}/SFR_{UV}$ is ill-defined (see Section\ \ref{obscuration.sec}) were excluded from the analysis.  For nearby galaxies, and even at $z \sim 1$, we find a surprisingly good agreement between model and observations.  A slight bias is present at $z \sim 1$, in the sense that the inferred $SFR_{IR}/SFR_{UV}$ for the bulk of the galaxies exceeds the observed values by a factor of $\sim 2$.  This systematic offset increases further toward higher redshift, amounting to an order of magnitude at $z \sim 2$.

We speculate briefly on the physical origin of this trend.  A refined model should account for an increased amount of UV emission escaping unhindered from high-redshift galaxies.  This situation is reminiscent of Adelberger \& Steidel (2000), who found that for galaxies of a given SFR, at high redshift a larger proportion of that star formation is emitted in the UV.  Daddi et al. (2007) also discussed that $z \sim 2$ galaxies are more "transparent" to UV radiation than local galaxies with similar levels of star-forming activity.  The offsets found in the $z \sim 1$ and $z \sim 2$ panels of Figure\ \ref{IRUVcomp.fig} could signal shortcomings in any of the three steps of our simple model.  Since the first two steps (star formation law, and metallicity scaling) are empirically motivated, and the third (geometry) is not, it makes sense to address the latter first.

Given that star formation, dust production, and the physical processes (such as winds) that determine the distribution of the ISM are intimately connected, the idealized model of a homogeneous mixture of dust and (young) stars will likely break down at some level.  If high-redshift galaxies are more patchy than their local counterparts, this would naturally explain why a relatively larger fraction of the intrinsic emission can escape the galaxies unattenuated.  In contrast to galaxies today, the UV and IR emission of high-redshift systems would in that case more often be associated with physically distinct regions within a galaxy.  If such a patchy configuration would indeed be more representative for $z \sim 2$ galaxies, we would expect this to be revealed in terms of rich internal color variations in multi-wavelength high-resolution imaging.  Such internal color variations, and their interpretation in terms of spatially resolved stellar populations will be the focus of a forthcoming paper (Wuyts et al. in prep).  First steps in this direction have been taken through visual classification (Cameron et al. 2010) and quantitative internal color dispersion methods (Bond et al. 2011) to compare ACS and WFC3 morphologies of distant galaxies in the ERS and HUDF regions.

Alternatively, the results of our model comparison may signal deviations from the adopted star formation law (Equation\ \ref{KS.eq}).  A tuning of the KS law to enforce an agreement between the modeled and observed $SFR_{IR}/SFR_{UV}$ would  require a higher normalization and/or a steeper slope than 1.17, so that a given $\Sigma_{SFR}$ corresponds to a smaller column of obscuring material.  Genzel et al. (2010) demonstrate that for a subsample of galaxies (those identified as mergers) a higher normalization is indeed appropriate.  Depending on their contribution to the overall galaxy population, this may lead to a somewhat steeper slope of the effective KS relation fitted to the combined sample of normal SFGs and mergers.  The change in slope required to bring our model prediction in agreement with the observations (keeping the zeropoint fixed) is substantial: a slope of 1.35 at $z \sim 1$ and 1.75 at $z \sim 2$.  KS slopes of this magnitude have been reported in the literature (e.g., Bouch\'{e} et al. 2007), but in combination with a lower zeropoint of the relation.  If instead we keep the slope fixed to 1.17, a zeropoint increase to -3.2 at $z \sim 1$ and -2.5 at $z \sim 2$ would be required to obtain agreement between model and observations.  A large fraction of $z \sim 2$ galaxies for which gas surface densities have been measured do lie on a relation with slope 1.17 and zeropoint -2.44, but that is because due to sensitivity limitations these samples have often been biased to high-SFR outliers.  Recent measurements for $z \sim 2$ galaxies on the main sequence (i.e., representative for the bulk of the $z \sim 2$ population) are consistent with the local relation for normal SFGs (Genzel et al. 2010; Daddi et al. 2010), and fitted by themselves give a zeropoint of -3.11 (and slope of 1.17).  The required change therefore seems to be in tension with the observational constraints at $z \sim 2$ available to date.

Finally, a lower dust column for a given $\Sigma_{SFR}$ could also be obtained if our metallicities of high-redshift galaxies are overestimated.  If this is the case, this would imply that gas-phase metallicity is determined by three parameters ($Z(SFR, M, z)$) rather than a non-evolving dependence on SFR and stellar mass.  We stress that current observations out to $z \sim 2.5$ do not require such an evolution with redshift of the SFR-Mass-Metallicity plane, but the high-redshift samples presently show a large scatter (Mannucci et al. 2010).  Moreover, while we are still facing large systematic uncertainties in metallicity measurements of even local galaxies, we already took the liberty to rescale the Mannucci et al. (2010) relation to reach a plateau of solar metallicity at the high-mass end, which is among the lower end of calibrations presented in the literature (Kewley et al. 2008).  Vice versa, if we were to adopt the Mannucci et al. (2010) relation without rescaling (i.e., reaching a plateau of $12 + \log (O/H) = 9.07$, or 2.6 times solar), the model prediction of $SFR_{IR}/SFR_{UV}$ would exceed the observed values by a factor 5, 10, and 60 at $z \sim 0.1$, $z \sim 1$, and $z \sim 2$.

We conclude that a simple model, combining a mixed dust-star geometry with an empirical star formation law and metallicity scaling, is successful in describing the obscuration and surface density of star formation in a self-consistent picture.  Towards higher redshifts, an adjustment of this model is desired, allowing more of the UV emission to escape.  This may be due to the fact that galaxies at higher lookback times feature more patchy geometries.

\section {Summary}
\label{summary.sec}

We use statistically significant samples of unprecedented size ($\sim 640000$ galaxies at $z \sim 0.1$, $\sim 130000$ at $z \sim 1$, and $\sim 36000$ at $z \sim 2$) to analyze how the structure and mode of star formation of galaxies at different lookback times depends on their position in the SFR-Mass diagram.  To this end, we combine 4 deep lookback surveys (COSMOS, UDS, GOODS-N, GOODS-S) with high-resolution imaging and rich multi-wavelength data sets reaching from the optical and near-infrared to mid- and far-infrared wavelengths.  Value-added SDSS catalogs are complemented with GALEX data to investigate equivalent galaxy properties in the nearby universe.

We find remarkable similarities over 10 Gyr of lookback time.  E.g., the relation between surface brightness profile shape on the one hand and the position of a galaxy with respect to the main sequence of star formation on the other hand is strikingly similar at all observed epochs.  Phrased differently, we find evidence for a correlation between galaxy structure and stellar populations (i.e., a Hubble sequence) to be in place already three billion years after the Big Bang.  The fact that a structurally distinct quiescent population below the main sequence is already present since $z \sim 2.5$ implies that the quenching mechanism responsible for the shutdown of star formation must be universally present already since that epoch, and must be causally connected to the morphological transition as well (or vice versa, see Martig et al. 2009 who analyze a morphological quenching scenario).

Low- and high-redshift galaxies also seem to follow a qualitatively similar behavior in terms of star formation surface density and its dependence on the specific SFR (see also Elbaz et al. 2011), and the variation in obscuration (parametrized by $SFR_{IR}/SFR_{UV}$) along and across the MS.  They differ in the sense that the high end of the relation between $\Sigma_{SFR}$ and $SFR/M$ is more abundantly populated at larger lookback times.

Another aspect in which low- and high-redshift galaxies differ is their typical size.  Measured at fixed $[SFR, M]$, galaxies on the MS today experienced a slower growth over time ($\sim (1+z)^{-0.4}$) than the quiescent population ($\sim (1+z)^{-1.2}$).  These growth rates are consistent with earlier work based on ground-based imaging (Trujillo et al. 2006; Franx et al. 2008; Toft et al. 2009) and recent WFC3 measurements by Nagy et al. (2011).\footnote{Nagy et al. (2011) find similar $r_e(z)/r_e(z=0)$ as star-forming galaxies in our sample that satisfy the same selection criteria.  Their steeper growth rate ($\sim (1+z)^{-1.42}$ within $1.5 < z < 3$) stems from the fact that they do not anchor the size evolution to $z=0$, i.e., they effectively use a different parametrization.}  A simple mass-size relation is inadequate to describe the size variations of the entire galaxy population.  At all epochs, star-forming galaxies on the MS are the largest among galaxies of a given mass.  

The profile shape of galaxies, quantified by the Sersic index $n$, is not a monotonic function of the level of star formation.  Instead, at a given stellar mass a reversal towards higher $n$ is observed in the high-SFR tail of the distribution.  Both our low- and high-redshift samples show evidence of such a relation.  This trend may hint at an evolutionary connection between the high-SFR outliers, which are in the process of building up a nuclear concentration of mass, and the quiescent galaxy population, which is characterized by cuspy surface brightness profiles, as proposed by the merger paradigm.

The variation of $\Sigma_{SFR}$ across the SFR-Mass diagram shows a different behavior than that of $SFR_{IR}/SFR_{UV}$.  While iso-$\Sigma_{SFR}$ regions run diagonal, following approximately lines of constant specific SFR (see also Schiminovich et al. 2007 for $z \sim 0.1$), the $SFR_{IR}/SFR_{UV}$ ratio increases as we move up along the MS (i.e., at fixed specific SFR).  This strongly suggests variations in metallicity along the main sequence, as expected from studies of the mass-metallicity (Erb et al. 2006; Buschkamp et al. in prep) and SFR-mass-metallicity (Mannucci et al. 2010) relation.  An increase in gas-phase metallicity towards the high-mass end of the MS would cause the obscuring column to increase, even if the gas column remains the same.

Using a simple model involving the observed Kennicutt-Schmidt star formation law (Genzel et al. 2010) and an empirically calibrated metallicity scaling (Mannucci et al. 2010) in combination with an assumed dust-star geometry, we compare the observed $SFR_{IR}/SFR_{UV}$ ratios to those inferred from the observed $\Sigma_{SFR}$.  Our model is successful at qualitatively relating the two properties, and explaining why their iso-contours in the SFR-Mass plane are of different shape.  For a metallicity scaling that flattens at solar values at the high-mass end, we find a good quantitative correspondence at $z \sim 0.1$ and $z \sim 1$ when assuming a homogeneous mixture of dust and stars.  At higher redshifts, progressively more patchy dust-star geometries appear to be required in order to have sufficient UV emission escaping to match the observed $SFR_{IR}/SFR_{UV}$.

\vspace{0.1in}

This work is based on observations taken by the CANDELS Multi-Cycle Treasury Program with the NASA/ESA HST, which is operated by the Association of Universities for Research in Astronomy, Inc., under NASA contract NAS5-26555.

Funding for the Sloan Digital Sky Survey (SDSS) has been provided by the Alfred P. Sloan Foundation, the Participating Institutions, the National Aeronautics and Space Administration, the National Science Foundation, the U.S. Department of Energy, the Japanese Monbukagakusho, and the Max Planck Society. The SDSS Web site is http://www.sdss.org/.

The SDSS is managed by the Astrophysical Research Consortium (ARC) for the Participating Institutions. The Participating Institutions are The University of Chicago, Fermilab, the Institute for Advanced Study, the Japan Participation Group, The Johns Hopkins University, Los Alamos National Laboratory, the Max-Planck-Institute for Astronomy (MPIA), the Max-Planck-Institute for Astrophysics (MPA), New Mexico State University, University of Pittsburgh, Princeton University, the United States Naval Observatory, and the University of Washington.

The Galaxy Evolution Explorer (GALEX) is a NASA Small Explorer. The mission was developed in cooperation with the Centre National d'Etudes Spatiales of France and the Korean Ministry of Science and Technology.


\onecolumngrid

\begin {figure*}[htbp]
\centering
\epsscale{0.8}
\plotone{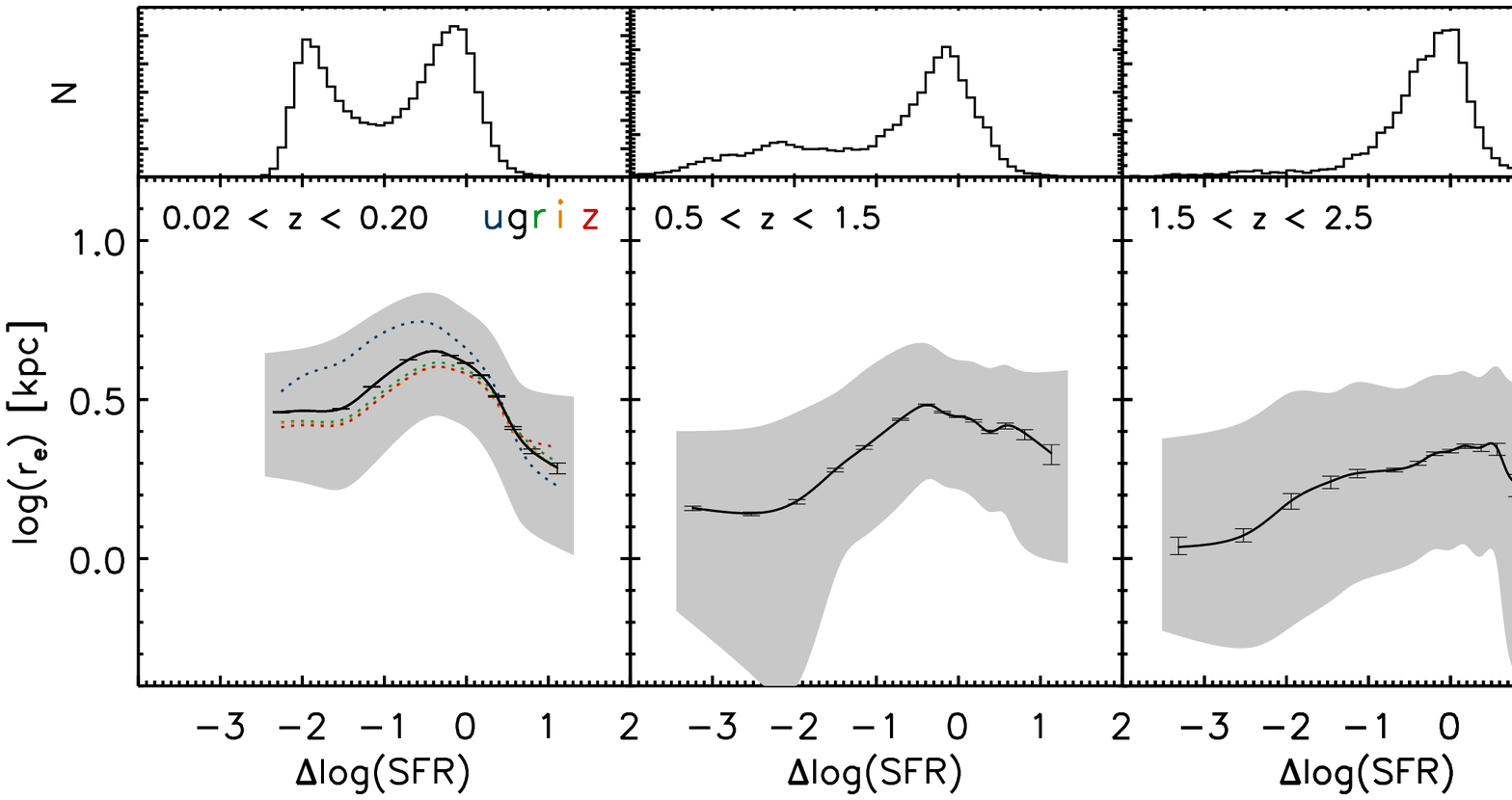} 
\plotone{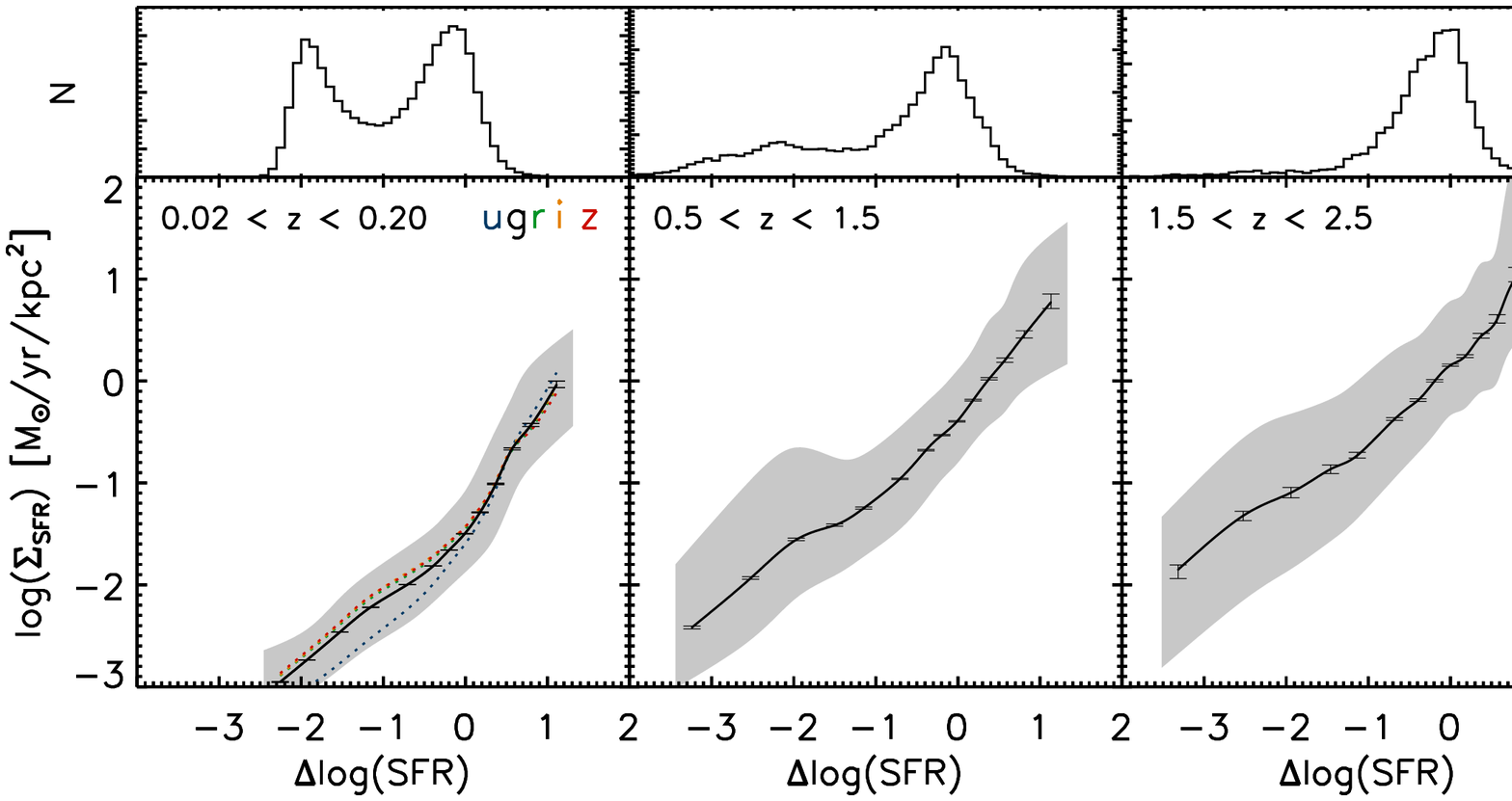} 
\plotone{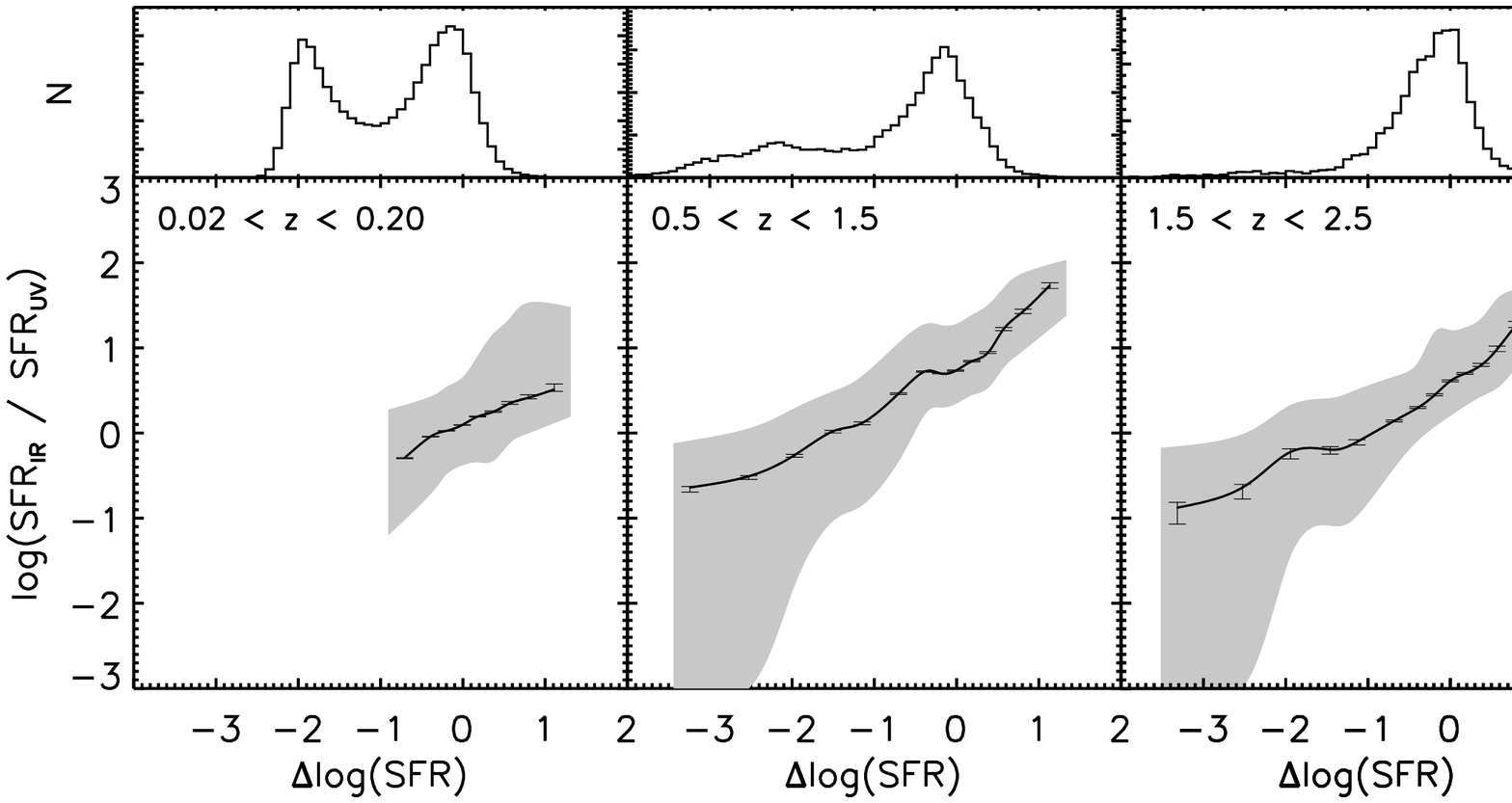} 
\epsscale{1.0}
\caption{
Idem as Figure\ \ref{dev.fig}, but for $\log(r_e)$, $\log(\Sigma_{SFR})$, and $\log(SFR_{IR}/SFR_{UV})$ instead of Sersic index.
\label{dev_app.fig}}
\end {figure*}

\vspace{0.2in}
\begin{center} APPENDIX A  GALAXY PROPERTIES ACROSS THE MAIN SEQUENCE\end{center}
Similar to Figure\ \ref{dev.fig} in Section\ \ref{profile.sec}, we here present the dependence of the effective radius $r_e$, surface density of star formation $\Sigma_{SFR}$, and obscuration $SFR_{IR}/SFR_{UV}$ of galaxies with $9.7 < \log M < 11$ as a function of deviation from the MS (Figure\ \ref{dev_app.fig}).  The scatter at a given $\Delta \log(SFR)$ ({\it grey polygons}) stems from a combination of uncertainties in the derived parameters and intrinsic variations within the galaxy population of a given specific SFR.  Especially the relations reflecting the mode of star formation ($\Sigma_{SFR}$ and $SFR_{IR}/SFR_{UV}$) with respect to the MS are remarkably tight.

\vspace{0.2in}
\begin{center} APPENDIX B  UNCERTAINTIES AND COMPLETENESS\end{center}

We follow two approaches to address the robustness of the trends in SFR-Mass space described in this paper.  First, we compare empirically the results for individual fields (B.1) and for data sets of different depth and wavelength in a given field (B.2).  Second, we use Monte Carlo simulations (B.3) to propagate different sources of uncertainty (photometry, redshifts, SFRs, structural parameters, ...) and assess their impact on the median properties of the overall galaxy population.  While specific measurements on individual objects may sometimes be subject to significant uncertainties, the conclusions drawn in this paper are much less affected as they are based on median properties over large numbers of galaxies, and the level of systematic biases is small.

\begin{center} B.1 Field-to-Field Variations and Completeness\end{center}

\begin{deluxetable*}{lrrrrrr}[t]
\tablecolumns{7}
\tablewidth{0pc}
\tablecaption{Overview Deep Lookback Surveys}
\tablehead{
\colhead{Field} & \colhead{Area} & \colhead{$Filter_{morph}$} & \colhead{Image Depth\tablenotemark{a}} & \colhead{Sample Depth\tablenotemark{b}} & \colhead{$N_{0.5 < z < 1.5}$\tablenotemark{c}} & \colhead{$N_{1.5 < z < 2.5}$\tablenotemark{c}} \\
\colhead{}      & \colhead{(deg$^2$)} & \colhead{}          & \colhead{(AB mag, 5$\sigma$)} & \colhead{(AB mag)} & \colhead{} & \colhead{}
}
\startdata
COSMOS          & 1.480                       & $I_{814}$   & 27.2           & 25.0        & 106080 & 21430 \\
UDS             & 0.056                       & $H_{160}$   & 26.7              & 26.7    & 10443 & 6796 \\
GOODS-S         & 0.041                       & $H_{160}$   & 27.0         & 27.0          & 7008 & 3973 \\
GOODS-N         & 0.042                       & $z_{850}$   & 27.6         & 26.8          & 8797 & 3450 \\
\enddata
\tablenotetext{a}{Point source depth of the image on which the morphological analysis was performed.}
\tablenotetext{b}{Magnitude (in $i$, $H_{160}$, $H_{160}$ and $z_{850}$ for COSMOS, UDS, GOODS-S and GOODS-N respectively) down to which galaxies were included in our sample.}
\tablenotetext{c}{Sample size in the $0.5 < z < 1.5$ and $1.5 < z < 2.5$ redshift intervals.}
\label{fields.tab}
\end{deluxetable*}

Our results at intermediate and high redshift are based on 4 lookback surveys, each with a wealth of multi-wavelength data.  They complement each other in area and depth, and differ in some cases in the selection band of the multi-wavelength catalog as well as the resolution and waveband of available HST imaging (see Table\ \ref{fields.tab}).  A comparison of the results obtained for each of the fields individually therefore serves as an empirical robustness check that encompasses field-to-field variations, the impact of imaging depth on morphological measurements and completeness within a given $[SFR, M]$ bin, resolution, and morphological k-corrections, if present.  In Figure\ \ref{field_per_field.fig}, we demonstrate that the median behavior of $n$, $r_e$, $\Sigma_{SFR}$ and $SFR_{IR}/SFR_{UV}$ looks strikingly similar for our samples in COSMOS, UDS, GOODS-S and GOODS-N, despite inhomogeneities between the data sets.  In cases where the patterns look somewhat less smooth, this is to be expected from the smaller number statistics.  From the similarity of the $\Sigma_{SFR}$ and $SFR_{IR}/SFR_{UV}$ diagrams, it follows that also our simple model and interpretation in terms of patchier dust geometries at high redshift holds for each of the fields separately.

\begin {figure*}[htbp]
\centering
\plotone{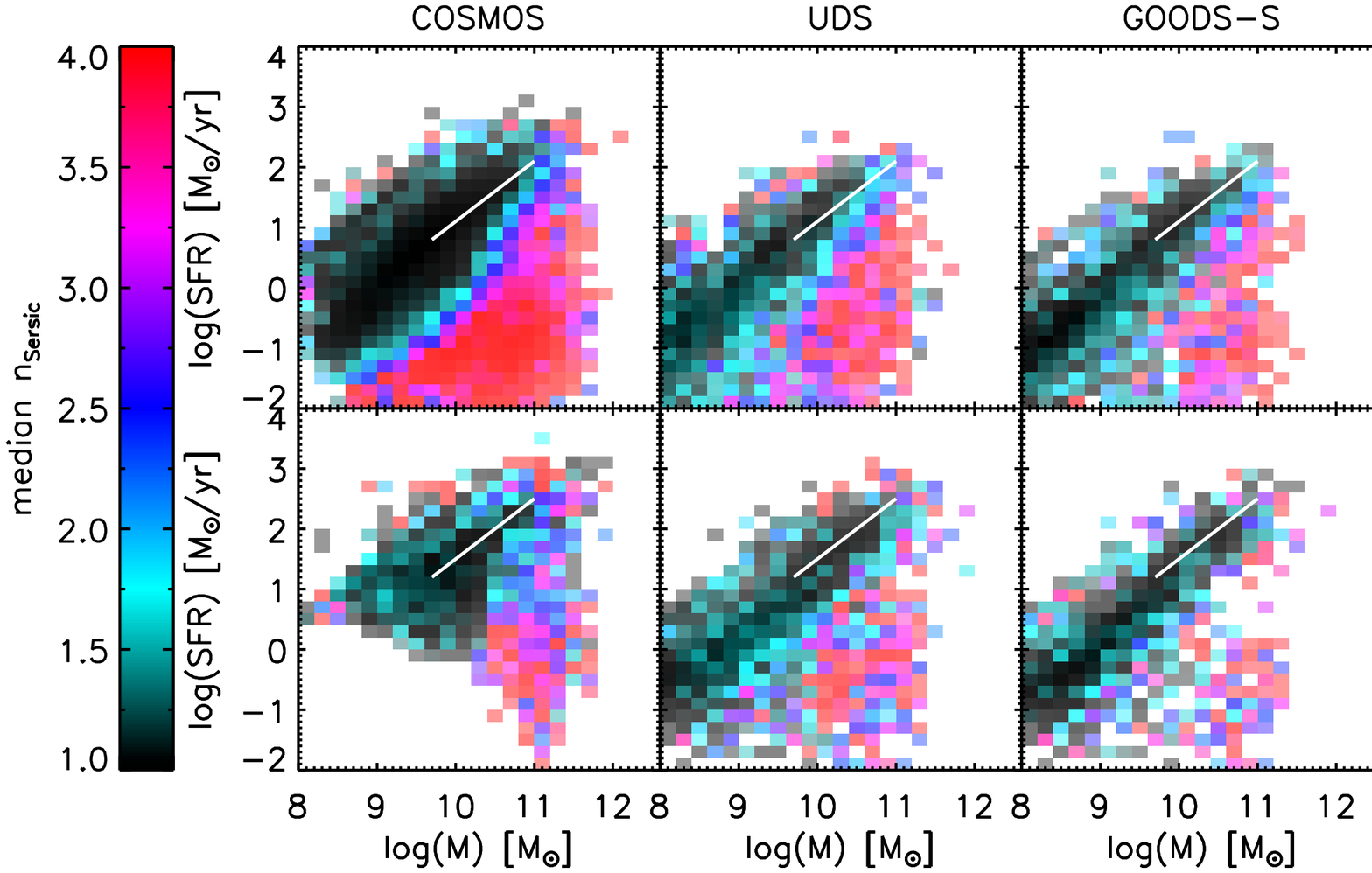}
\vspace{0.2in}
\plotone{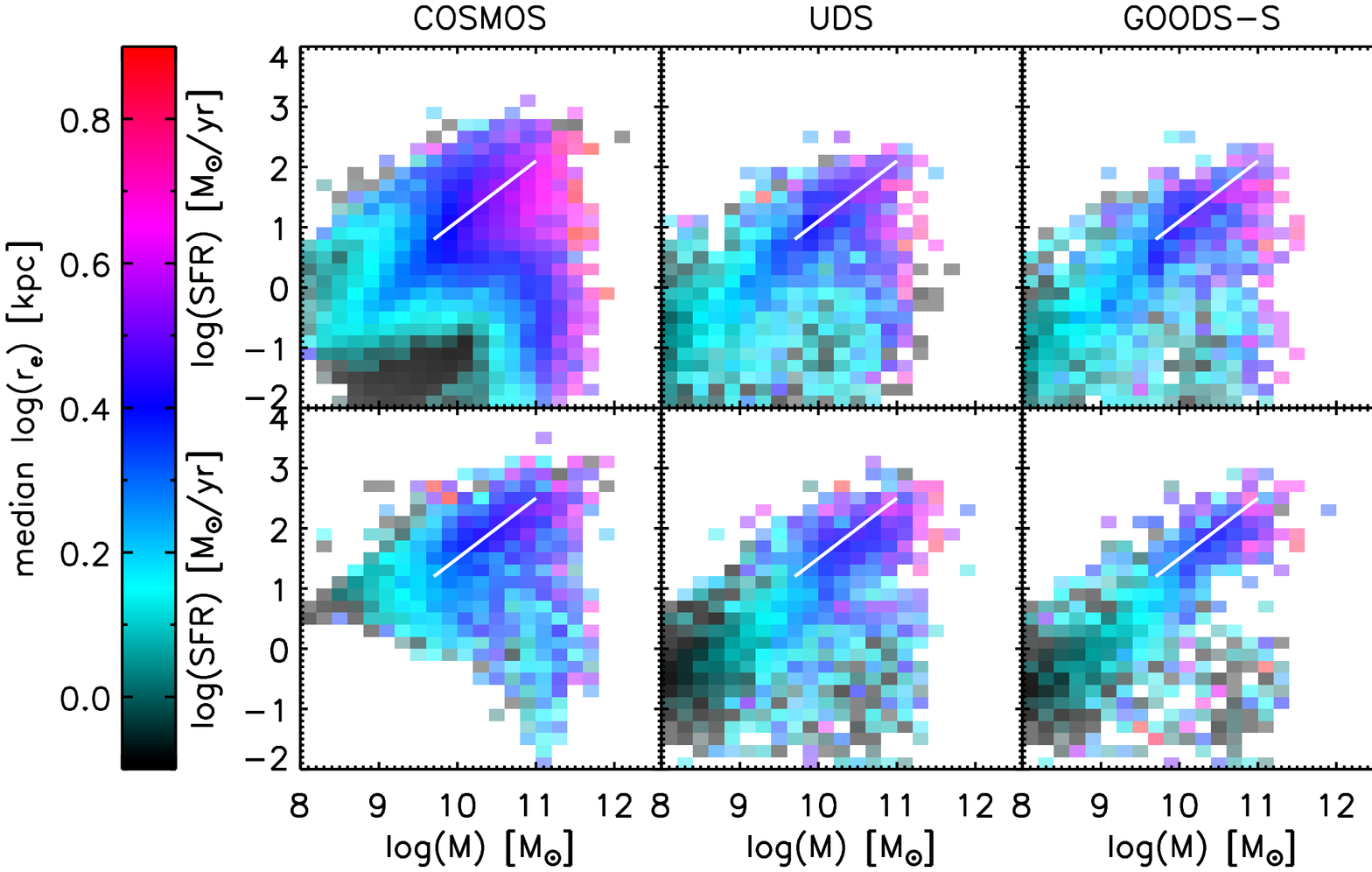}
\caption{
{\it From top to bottom:} SFR versus mass diagrams of the 4 deep fields separately, color-coded by median Sersic index, size, surface density of star formation, and $SFR_{IR}/SFR_{UV}$ ratio.  Despite differences in depth, and waveband used to perform the morphological analysis, the patterns in SFR-mass space are consistent in all fields, boosting confidence in the robustness of our results.
\label{field_per_field.fig}}
\end {figure*}

\setcounter{figure}{8}
\begin {figure*}[htbp]
\plotone{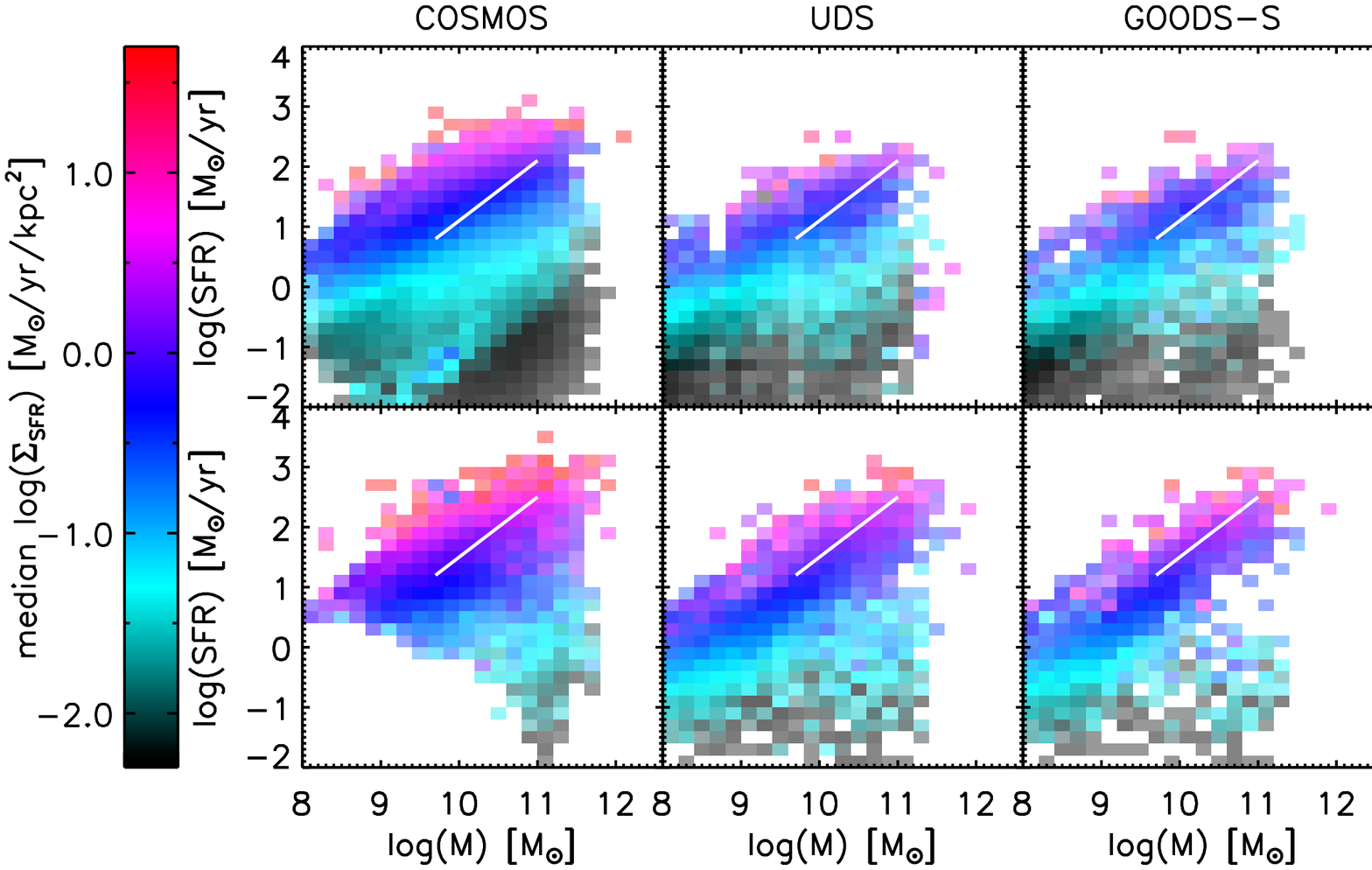}
\vspace{0.2in}
\plotone{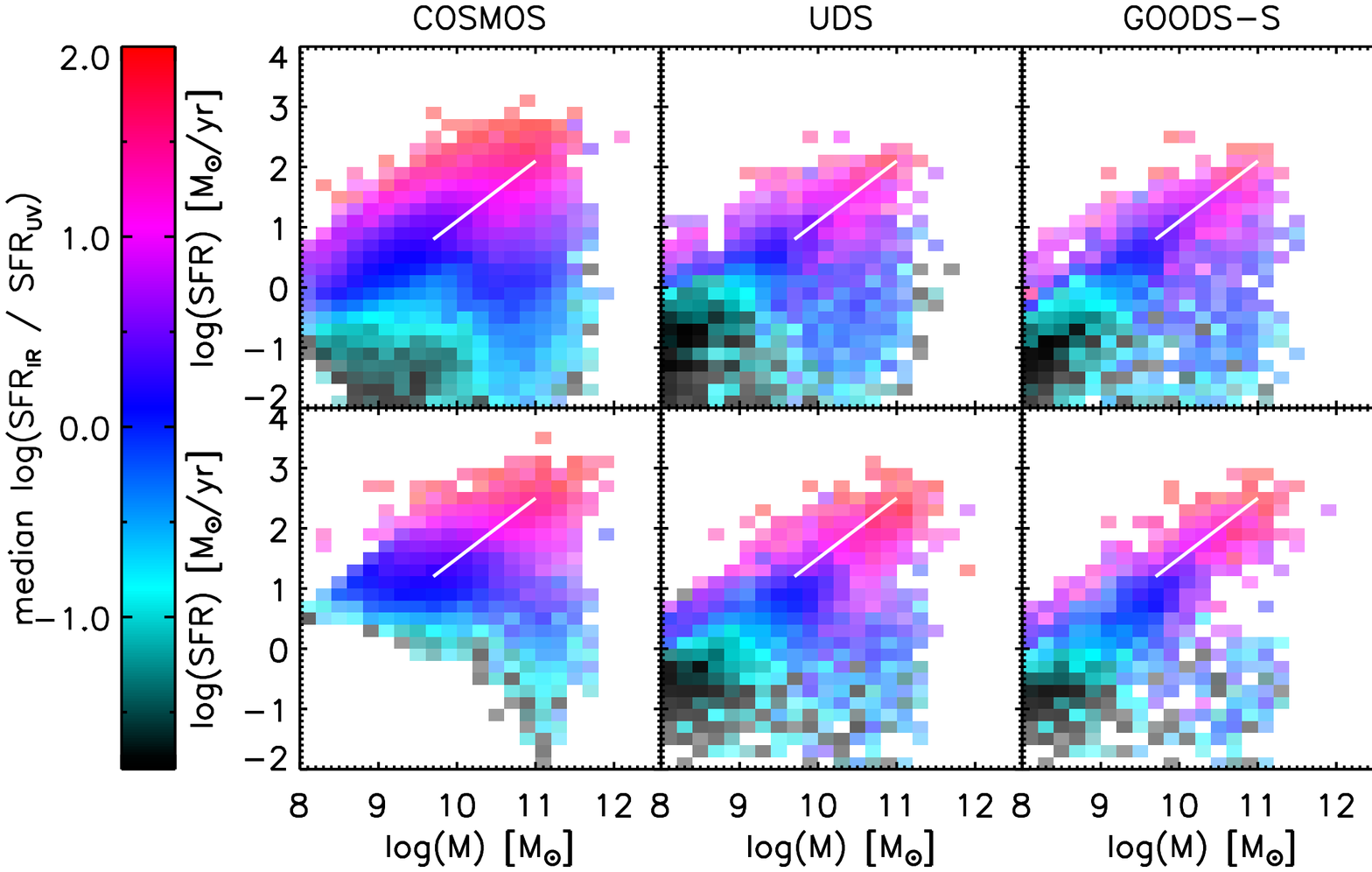}
\caption{
continued
\label{field_per_field.fig}}
\end {figure*}

\begin {figure*}[htbp]
\centering
\plotone{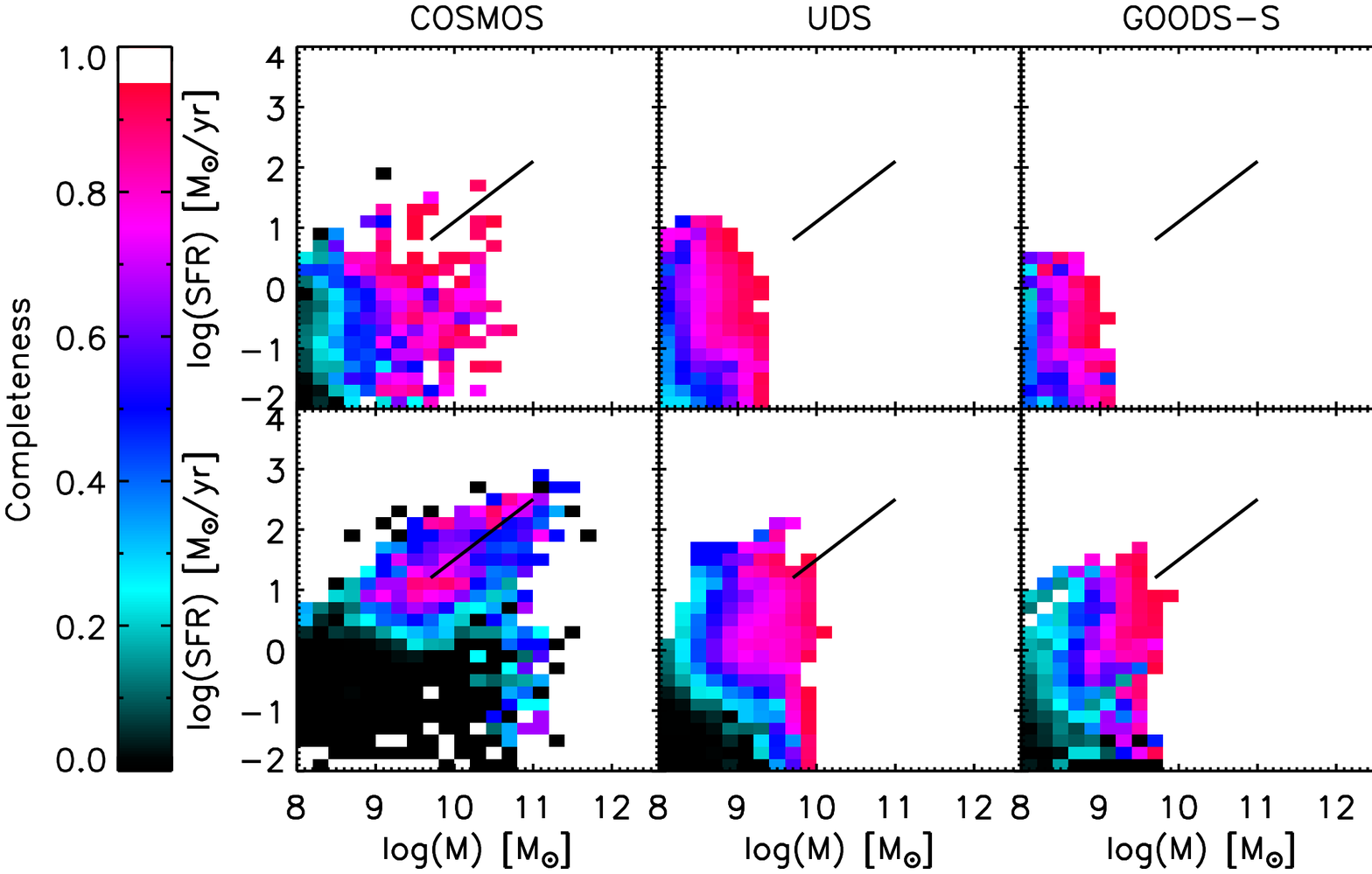}
\caption{
Completeness of our high-redshift samples in COSMOS, UDS, GOODS-South, and GOODS-North as a function of SFR and mass.
\label{completeness.fig}}
\end {figure*}

The region of SFR-Mass space covered by each of the fields in Figure\ \ref{field_per_field.fig}, reflects differences in depth.  We expand on this in Figure\ \ref{completeness.fig}, where the color-coding illustrates the estimated completeness of our sample.  For the COSMOS and GOODS-N fields, the completeness in each $[SFR, M]$ bin was defined empirically as the fraction of UDS and GOODS-S galaxies that would enter the $i$- and $z$-band limited samples for those fields respectively.  For the deep WFC3 $H$-band selected samples in UDS and GOODS-S, we estimated the completeness indirectly.  For a given $[SFR, M]$ bin, we dimmed the SEDs of galaxies with similar specific SFR, but higher SFR and mass (i.e., located in a highly complete region of the diagram).  As completeness estimate, we adopted the fraction of galaxies downscaled to the $[SFR, M]$ bin of interest that would still be sufficiently bright to enter the sample.  The latter is best interpreted as an upper bound on the completeness.  As extinction is known to scale with the star formation activity (see, e.g., the increased obscuration as one moves along the MS toward the high-mass end in Figure\ \ref{IRUVobs.fig}), our simple approach of downscaling the SED by the same factor as the SFR and mass, without altering the SED shape, likely leads to a completeness estimate that is too conservative.

Figure\ \ref{completeness.fig} illustrates that the completeness level of our COSMOS sample is lower than that of the samples extracted from the other fields.  As is evident from Figure\ \ref{field_per_field.fig}, COSMOS therefore gets zero weight at the lowest SFRs and masses probed in the $z \sim 2$ bin.  In $[SFR, M]$ bins that do contain COSMOS galaxies, two tests indicate that the somewhat lower completeness level does not lead to significant biases in the median morphology or mode of star formation.  First, the trends in COSMOS do not differ from those observed in the other fields (Figure\ \ref{field_per_field.fig}).  Second, we applied the same $i$-band limit adopted for COSMOS to the highly complete WFC3-selected catalog in UDS.  We verified that this does not change the median morphological and mode of star formation parameters in $[SFR, M]$ bins that still contain sources after imposing the $i$-band cut.

\begin{center} B.2 Depth and Wavelength Dependence of Morphological Measurements \end{center}

\begin {figure*}[htbp]
\centering
\plottwo{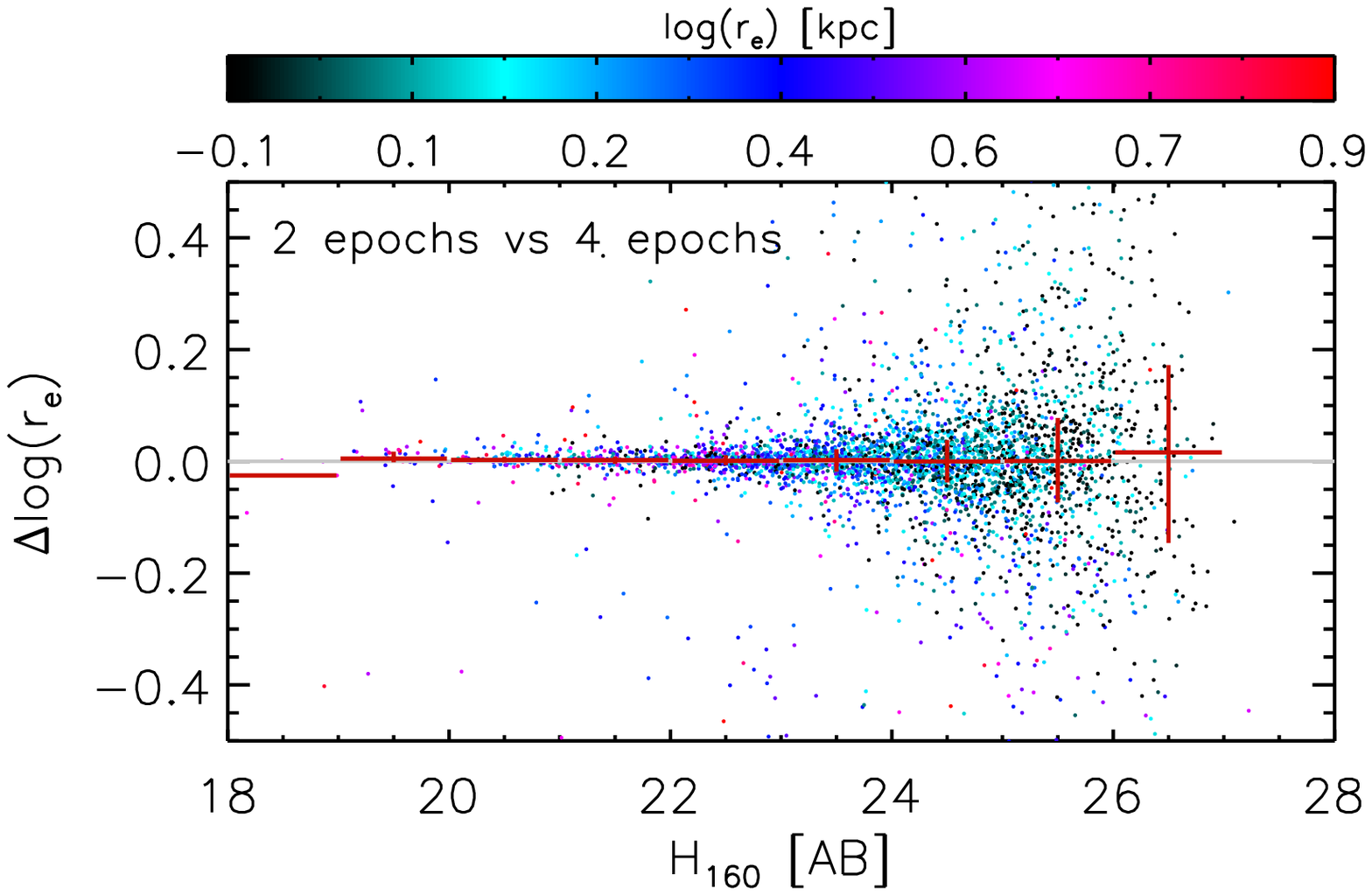}{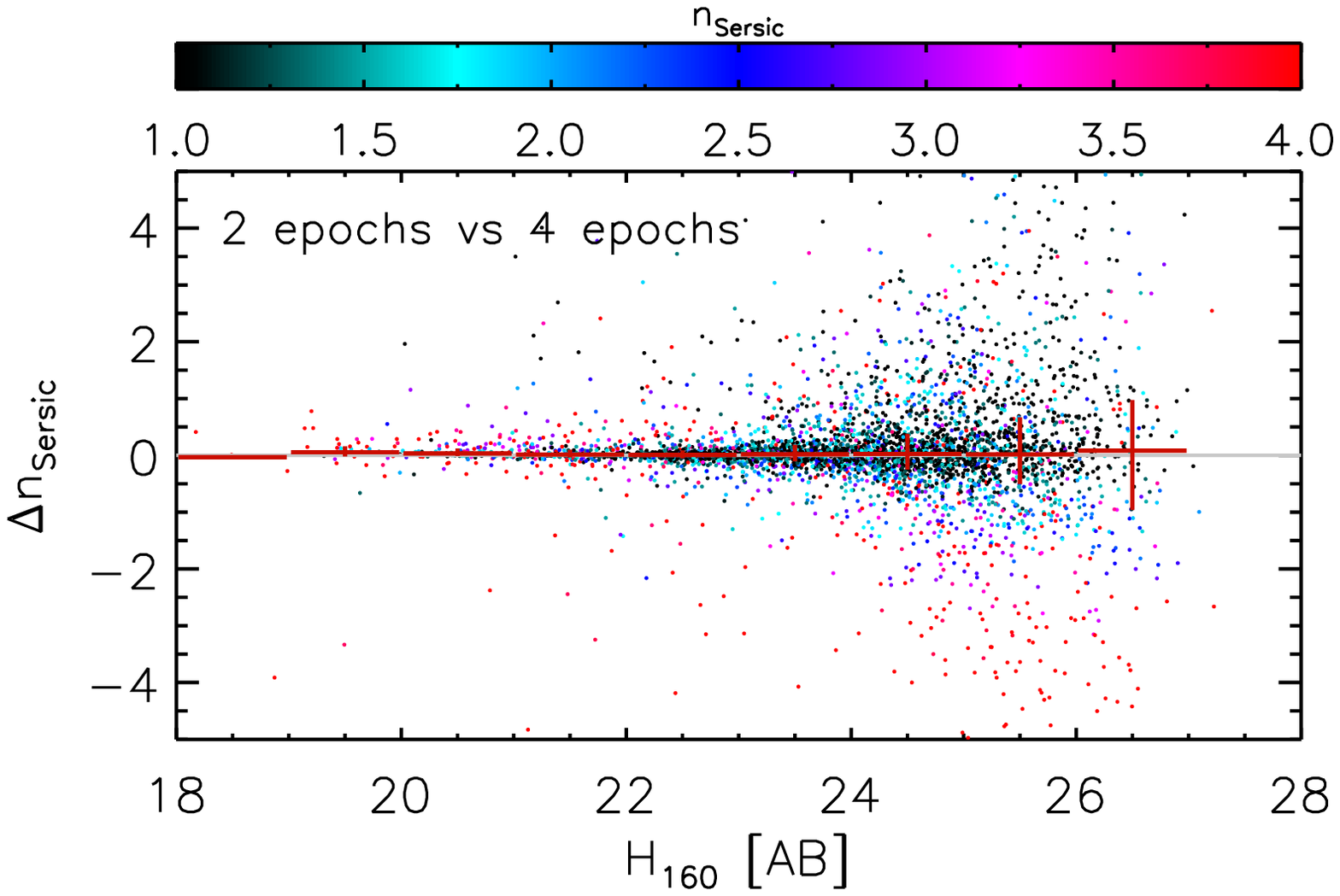}
\plottwo{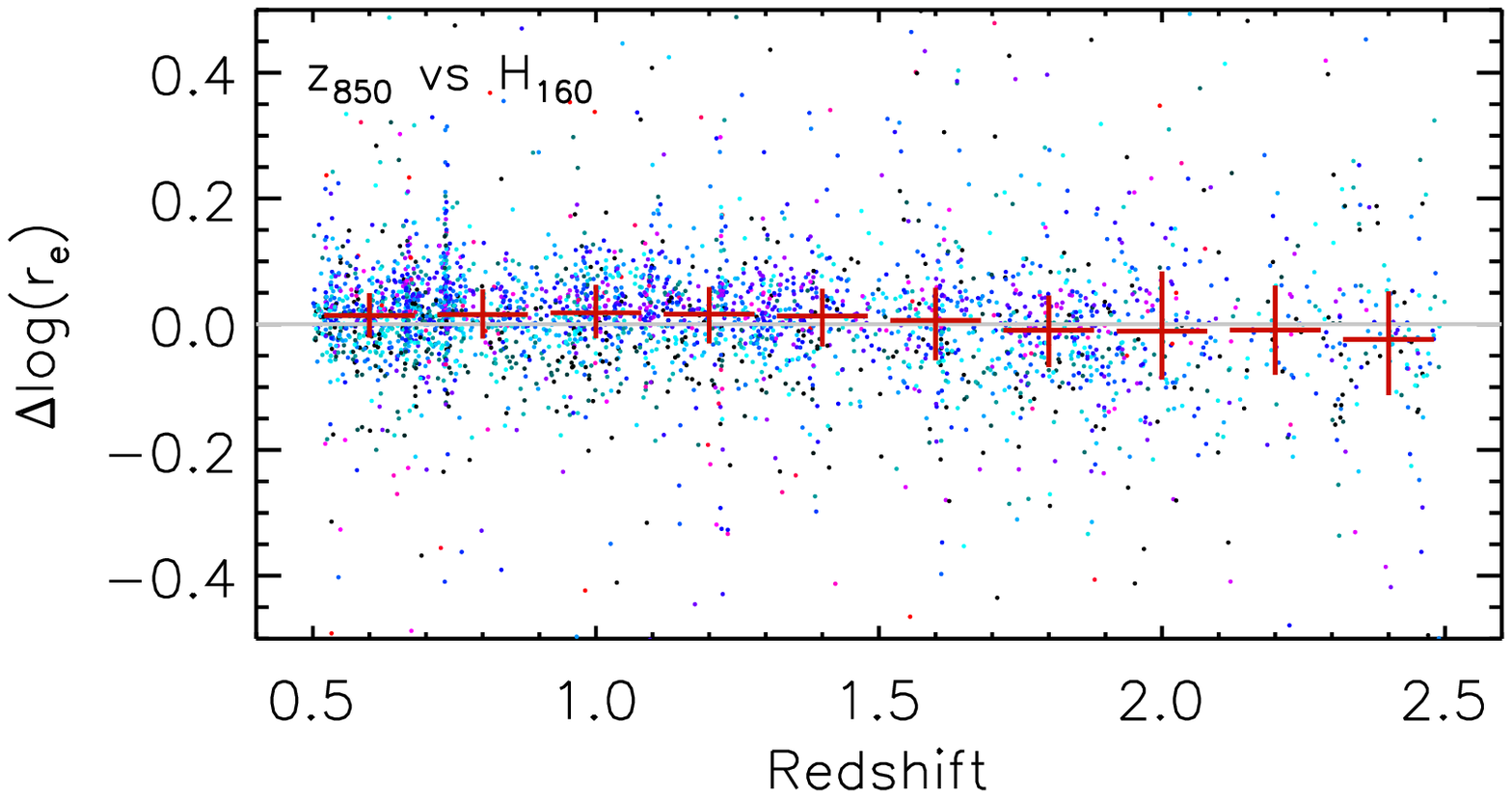}{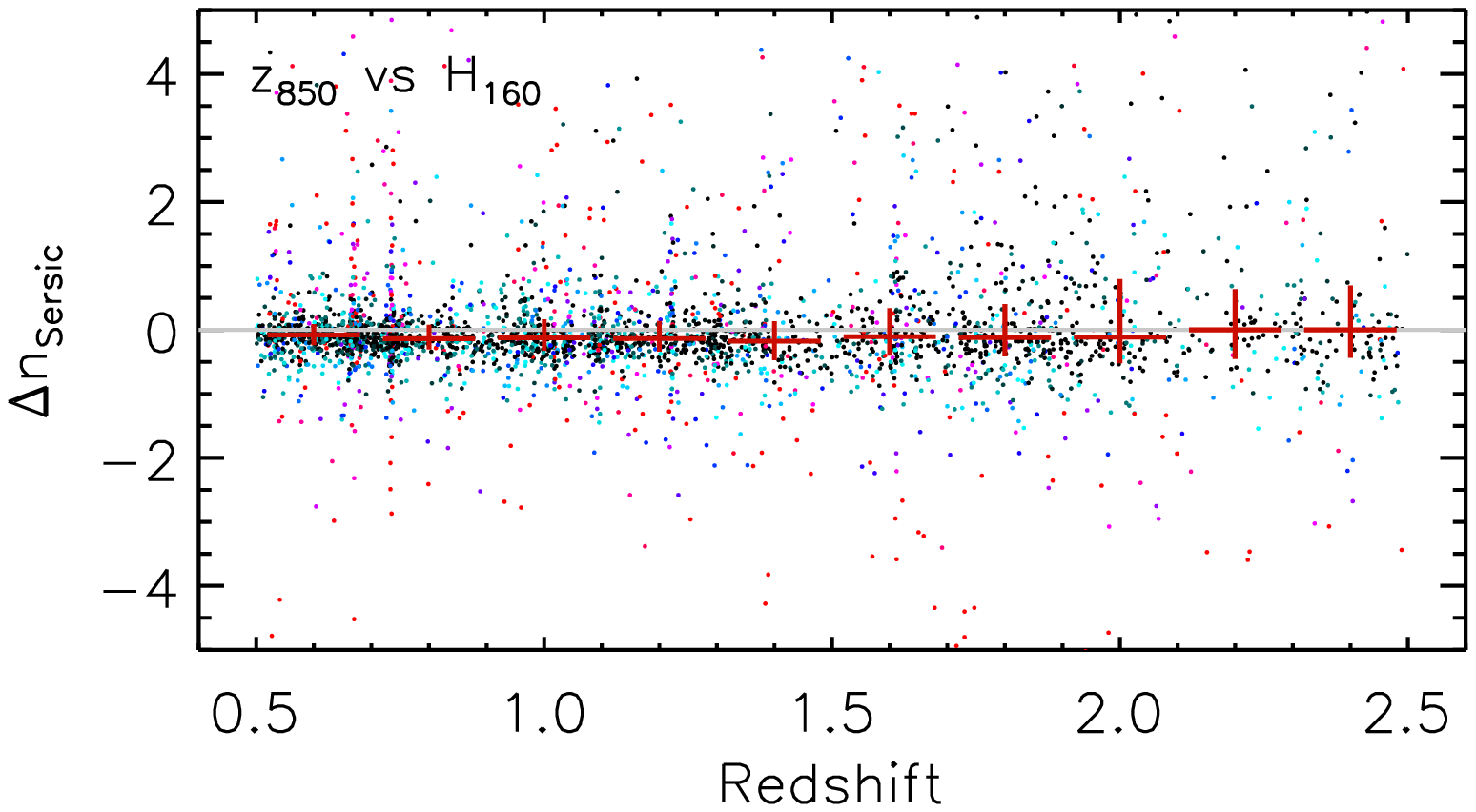}
\caption{
Comparison between different GALFIT runs in GOODS-South.  {\it Top panels:} Deviation in half-light radius and Sersic index measured on WFC3 $H_{160}$ mosaics of 2 epochs versus 4 epochs depth as a function of $H_{160}$ magnitude. {\it Bottom panels:} Deviation in half-light radius and Sersic index measured on ACS $z_{850}$ versus WFC3 $H_{160}$ imaging as a function of redshift.  Horizontal red lines indicate the running median, while vertical red lines mark the central 50th percentile.  Systematic deviations from zero are mostly limited to a few percent, much smaller than the variation of morphological parameters across the SFR-Mass plane.  Offsets in Sersic index at the 10\% level are present between $z_{850}$ and $H_{160}$, such that the surface brightness profiles are slightly cuspier in the longer wavelength band.
\label{depth_wave.fig}}
\end {figure*}

While boosting confidence in the robustness of our results, the field-to-field comparison in B.1 did not discriminate between different aspects of inhomogeneity between the data sets.  Using the GOODS-S field as a test case, we now isolate the effect of depth, and the effect of waveband on which the surface brightness profile fitting is performed.  The top panels of Figure\ \ref{depth_wave.fig} illustrate how the morphological parameters change when measured on the presently available full depth mosaic of the CANDELS-Deep region, and on a version of the mosaic built with only the first half of the exposures (0.35 mag shallower, similar in depth to the CANDELS-Wide data used for UDS).  Unsurprisingly, the scatter in $\Delta \log (r_e)$ is observed to increase toward fainter magnitudes.  However, down to the faintest magnitudes we find no evidence for a systematic bias in the size measurement.  This implies that population studies which describe the median size, such as presented in this paper, can safely probe to fainter levels than would be considered reliable on an individual object level.  H\"{a}ussler et al. (2007) arrive at the same qualitative conclusion when inserting simulated $n=1$ and $n=4$ galaxies in mock ACS images, and recovering the morphological parameters using GALFIT.  For similar tests aimed specifically at the CANDELS data sets, we defer the reader to van der Wel et al. (in prep).  It is relevant to note that our conclusion may not necessarily hold for other methods to determine $r_e$ than the GALFIT code used in our analysis (see also H\"{a}ussler et al. 2007; Sargent et al. 2007).

The difference $\Delta n$ between the Sersic index measured on the 2- and 4-epoch mosaics also exhibits an increased scatter toward fainter magnitudes, but again the median $\Delta n$ remains consistent with zero.  In detail, however, a trend is noted such that shallow profile shapes are on average recovered correctly down to the faintest magnitudes, while the Sersic index of intrinsically cuspy systems shows a systematic deviation at the faint end.\footnote{We note that deviations $\Delta n$ of order -2 or lower can by definition not be reached for disk-dominated systems because of the allowed range $0.2 < n < 8$.}  In relative terms, the underestimate for faint $n > 4$ galaxies corresponds to a median $\Delta n / n \approx -0.4$.  In SFR-Mass space, these galaxies are predominantly located at $\log(M)<9$ and $\log(SFR)<0.5$.  If they were the dominant population in that area of SFR-Mass space, they could potentially bias the median $n$ to low values.  Our deepest, 4 epoch GOODS-S data suggests however that they only account for less than 10\% of the galaxy population in the low mass and low SFR regime.

The bottom panels of Figure\ \ref{depth_wave.fig} compare measurements of $n$ and $r_e$ carried out on the WFC3 $H_{160}$ mosaic, and on the ACS $z_{850}$ mosaic.  We plot the deviation (negative where the $z_{850}$ measurement yields a smaller value) as a function of redshift to facilitate the interpretation in terms of morphological k-corrections in the rest-frame.  Variations in the extent of the $z_{850}$- and $H_{160}$-band surface brightness profiles are small compared to the amplitude of the size variation across the SFR-Mass diagram.  Median offsets are limited to the few percent level for $r_e$, and of order $\sim 10$\%  for the Sersic index (such that the longer wavelength surface brightness profiles tend to be cuspier).  Overall, the systematic deviations in $n$ are small compared to the amplitude of trends with profile shape across the SFR-Mass plane, although we caution that the random scatter in $n$ due to morphological k-corrections becomes substantial for systems with $n > 4$ in the $H_{160}$ band.

\begin{center} B.3 Propagating Uncertainties with Monte Carlo Simulations\end{center}

\begin {figure*}[t]
\centering
\plotone{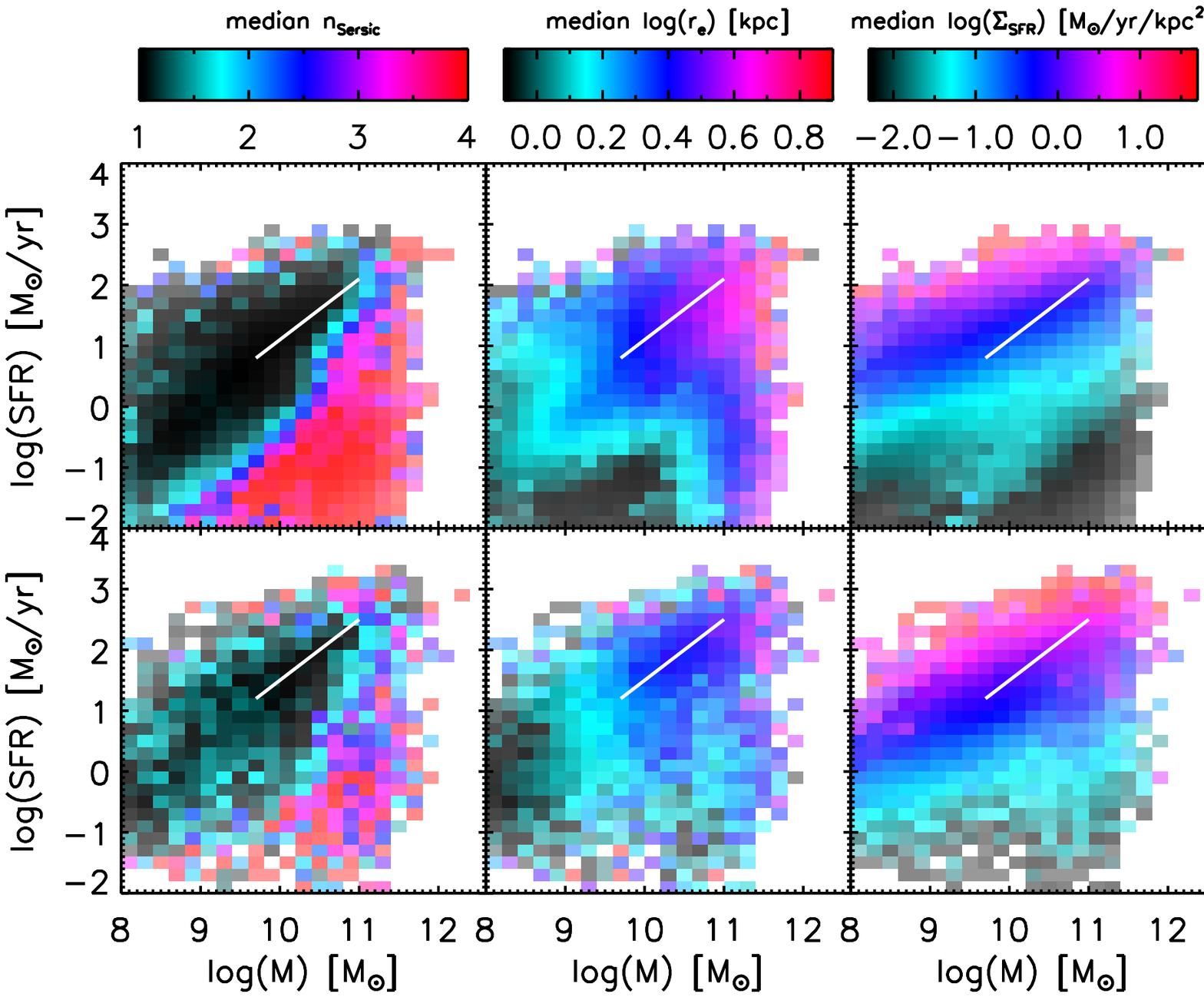}
\caption{
Monte Carlo simulated SFR vs mass diagrams at $z \sim 1$ and $z \sim 2$, constructed by perturbing the measured photometry and photometric redshifts according to their uncertainties and propagating to the other parameters of interest.  The characteristic patterns in SFR-Mass space remain robustly present in the MC simulation.
\label{MC.fig}}
\end{figure*}

We ran extensive Monte Carlo (MC) simulations to estimate the effects of photometric and photometric redshift uncertainties on our analysis.  First, we selected a representative subset of galaxies in each of the four deep fields.  We sample four redshift intervals ($0.5 < z < 1.0$, $1.0 < z < 1.5$, $1.5 < z < 2.0$, and $2.0 < z < 2.5$), drawing for each one three galaxies from each $[SFR, M]$ bin, spanning the lower, middle and upper tertile in signal-to-noise.  For the resulting subsample of 7480 galaxies, we then compiled one hundred mock photometric catalogs.  The flux-points in each optical-to-IR band were drawn randomly from a Gaussian with the measured flux as the mean and its error as the standard deviation.  Next, we subject each mock catalog to the same fitting procedures to derive photometric redshifts, stellar masses, and SFRs, and also propagate the newly obtained redshifts into the conversion from angular to physical half-light radii.  Together, the 100 mock realizations provide us with an uncertainty distribution on (and often correlated between) each of the parameters that enter our analysis.

In order to apply the MC results, we associate each galaxy from our full sample with a galaxy in the MC subsample that lies in the same field, and occupies the same redshift, SFR, mass and signal-to-noise bin.  The deviations with respect to the actual measurements of redshift, mass, SFR, and $r_e$ are drawn from one of the 100 mock realizations of the associated MC target, and applied to the galaxy of interest.  The resulting mock population is then again divided in $[SFR, M]$ bins and we compute the median of the property of interest ($n$, $r_e$, $\Sigma_{SFR}$, $SFR_{IR}/SFR_{UV}$), exactly as we did for our analysis of the real measurements.  In principle, the deviations in derived parameters (which unlike the photometry are not assumed or forced to be symmetric) should be applied to error-free measurements.  Since the latter are not available, our test serves as a conservative estimate of how important photometric errors are in our analysis.

Upon visual inspection of the MC-ed equivalents of Figures\ \ref{n.fig} to\ \ref{IRUVcomp.fig}, we find all the trends described in this paper to be robust against photometric uncertainties.  However, the $z_{phot}$ uncertainty distribution as inferred from running EAZY on the 100 perturbed photometric catalogs, is likely too narrow.  We find evidence for this by running a similar set of 100 simulations on the full spectroscopically confirmed sample.  The spectroscopic redshift lies within the formal 1$\sigma$ confidence interval derived from the MC simulation in only about 30\% of the cases, much less than the expectation value of 68\% were photometric errors the only source of uncertainty in deriving the redshift.

In a second iteration, we therefore base deviations in $z_{phot}$ on the empirical comparison between photometric redshifts and the available spectroscopic redshifts.  Photometric and photometric redshift uncertainties are then propagated to the other parameters as before.  Here, it is important to account for the fact that the spectroscopic sample is not randomly drawn from the overall galaxy sample, but instead biased towards bright magnitudes.   Following Scarlata et al. (2007), we quantified the photometric redshift uncertainty at a given magnitude semi-empirically by dimming the SEDs of spectroscopically confirmed galaxies to that magnitude (perturbing the photometry accordingly), and deriving the resulting $\Delta z / (1+z)$ distribution.  As expected, this yields a broader distribution, at the faintest levels by up to a factor of 5 compared to the normalized median absolute deviation of the undimmed spectroscopic sample.  However, even at the faintest levels, no significant systematic offsets in $z_{phot}$ are revealed.  

Figure\ \ref{MC.fig} presents the MC-ed equivalents of Figures\ \ref{n.fig} to\ \ref{IRUVobs.fig}, where photometric errors and semi-empirical photometric redshift uncertainties have been applied and propagated in the analysis.  We considered the distribution in $\Delta \wp = \wp_{MC} - \wp_{true}$, where $\wp$ stands for the median structural ($n$, $r_e$) or mode of star formation ($\Sigma_{SFR}$, $SFR_{IR}/SFR_{UV}$) property in a given $[SFR, M]$ cell.  Systematic deviations from zero are limited to $\lesssim 0.03$ dex.  At $0.5 < z < 1.5$, the scatter of the $\Delta \wp$ distribution is modest, amounting to 0.04, 0.04, 0.07, and 0.12 dex in $n$, $r_e$, $\Sigma_{SFR}$ and $SFR_{IR}/SFR_{UV}$ respectively.  At $1.5 < z < 2.5$, the scatter in $\Delta \wp$ doubles.   We conclude that, when following this more conservative approach, we still find that the patterns in SFR-Mass space described in this paper are robust against photometric and photometric redshift uncertainties.

In a third and final iteration, we now also include random uncertainties in mass, $SFR$, $r_e$, and $n$ that do not stem from photometric or photometric redshift uncertainties.  To this end, we adopt fiducial errors of 0.2 dex in each of these parameters.  This crudely reflects our ability to recover the intrinsic surface brightness profile shape (B.2), as well as imperfections with which observables are translated to physical parameters (see, e.g., Wuyts et al. 2009, 2011a).  The scatter in $\Delta \wp$ is boosted by $\sim 50\%$ compared to the MC simulation that only accounted for photometric and photometric redshift uncertainties.  Nevertheless, the overall conclusions drawn in this paper remain valid.


\begin{references}
{\small
\reference{} Abazajian, K. N., et al. 2009, ApJS, 182, 543
\reference{} Adelberger, K. L.,\& Steidel, C. C. 2000, ApJ, 544, 218
\reference{} Alexander, D. M., et al. 2003, AJ, 126, 539
\reference{} Asplund, M., Grevesse, N., Sauval, A. J., Allende Prieto, C.,\& Kiselman, D. 2004, A\&A, 417, 751
\reference{} Barger, A. J., Cowie, L. L., Sanders, D. B., Fulton, E., Taniguchi, Y., Sato, Y., Kawara, K.,\& Okuda, H. 1998, Nature, 394, 248
\reference{} Barger, A. J., Cowie, L. L.,\& Wang, W.-H. 2008, ApJ, 689, 687
\reference{} Barnes, J. E.,\& Hernquist, L. 1991, ApJ, 370, 65
\reference{} Barnes, J. E.,\& Hernquist, L. 1996, ApJ, 471, 115
\reference{} Bell, E. F. 2008, ApJ, 682, 355
\reference{} Berta, S., et al. 2010, A\&A, 518, 30
\reference{} Bianchi, L., Efremova, B., Herald, J., Girardi, L., Zabot, A., Marigo, P.,\& Martin, C. 2011, MNRAS, 411, 2770
\reference{} Blain, A. W., Smail, I., Ivison, R. J., Kneib, J.-P.,\& Frayer, D. T. 2002, PhR, 369, 111
\reference{} Birnboim, Y.,\& Dekel, A. 2003, MNRAS, 345, 349
\reference{} Blanton, M. R., et al. 2003, ApJ, 594, 186
\reference{} Blanton, M. R., et al. 2005, AJ, 129, 2562
\reference{} Bohlin, R. C., Savage, B. D.,\& Drake, J. F. 1978, ApJ, 224, 132
\reference{} Bond, N. A., Gawiser, E.,\& Koekemoer, A. M. 2011, ApJ, 729, 48
\reference{} Bouch\'{e}, N., et al. 2007, ApJ, 671, 303
\reference{} Brammer, G. B., van Dokkum, P. G.,\& Coppi, P. 2008, ApJ, 686, 1503
\reference{} Brinchmann, J., Charlot, S., White, S. D. M., Tremonti, C., Kauffmann, G., Heckman, T.,\& Brinkmann, J. 2004, MNRAS, 351, 1151
\reference{} Bruzual, G,\& Charlot, S. 2003, MNRAS, 344, 1000
\reference{} Calzetti, D., Armus, L., Bohlin, R. C., Kinney, A. L., Koornneef, J.\& Storchi-Bergmann, T. 2000, ApJ, 533, 682
\reference{} Cameron, E., Carollo, C. M., Oesch, P. A., Bouwens, R. J., Illingworth, G. D., Trenti, M., Labb\'{e}, I.,\& Magee, D. 2010, submitted to ApJ (arXiv1007.2422)
\reference{} Cappelluti, N., et al. 2009, A\&A, 497, 635
\reference{} Cassata, P., et al. 2011, ApJ, in press (arXiv1106.4308)
\reference{} Chabrier, G. 2003, PASP, 115, 763
\reference{} Conroy, C.,\& Wechsler, R. H. 2009, ApJ, 696, 620
\reference{} Cox, T. J., Jonsson, P., Somerville, R. S., Primack, J. R.,\& Dekel, A. 2008, MNRAS, 384, 386
\reference{} Daddi, E., et al. 2005, ApJ, 626, 680
\reference{} Daddi, E., et al. 2007a, ApJ, 670 156
\reference{} Daddi, E., et al. 2010, ApJ, 714, 118
\reference{} Dekel, A.,\& Birnboim, Y. 2006, MNRAS, 368, 2
\reference{} Dekel, A., et al. 2009, Nature, 457, 451
\reference{} Di Matteo, T., Springel, V.,\& Hernquist, L. 2005, Nature, 433, 604
\reference{} Elbaz, D., et al. 2007, A\&A, 468, 33
\reference{} Elbaz, D., et al. 2010, A\&A, 518, 29
\reference{} Elbaz, D., et al. 2011, A\&A, 533, 119
\reference{} Erb, D. K., Shapley, A. E., Pettini, M., Steidel, C. C., Reddy, N. A.,\& Adelberger, K. L. 2006, ApJ, 644, 813
\reference{} F\"{o}rster Schreiber, N. M., Genzel, R., Lutz, D., Kunze, D.,\& Sternberg, A. 2001, ApJ, 552, 544
\reference{} F\"{o}rster Schreiber, N. M., et al. 2009, ApJ, 706, 1364
\reference{} F\"{o}rster Schreiber, N. M., et al. 2011, ApJ, 731, 65 
\reference{} Franx, M., van Dokkum, P. G., F\"{o}rster Schreiber, N. M., Wuyts, S., Labb\'{e}, I.,\& Toft, S. 2008, ApJ, 688, 770
\reference{} Gabasch, A., Goranova, Y., Hopp, U., Noll, S.,\& Pannella, M. 2008, MNRAS, 383, 1319
\reference{} Genzel, R., et al. 2008, ApJ, 687, 59
\reference{} Genzel, R., et al. 2010, MNRAS, 407, 2091
\reference{} Giavalisco, M., et al. 2004, ApJ, 600, 93
\reference{} Grogin, N. A., et al. 2011, submitted to ApJS (arXiv1105.3753)
\reference{} Guo, Y., et al. 2011, ApJ, 735, 18
\reference{} H\"{a}ussler, B., et al. 2007, ApJS, 172, 615
\reference{} Heckman, T. M., Robert, C., Leitherer, C., Garnett, D. R.,\& van der Rydt, F. 1998, ApJ, 503, 646
\reference{} Hopkins, A. M.,\& Beacom, J. F. 2006, ApJ, 651, 142
\reference{} Hopkins, P. F., Hernquist, L., Cox, T. J., Di Matteo, T., Robertson, B.,\& Springel, V. 2006, ApJS, 163, 1
\reference{} Hopkins, P. F., Hernquist, L., Cox, T. J., Keres, D.,\& Wuyts, S. 2009, ApJ, 691, 1424
\reference{} Hopkins, P. F., Younger, J. D., Hayward, C. C., Narayanan, D.,\& Hernquist, L. 2010, MNRAS, 402, 1693
\reference{} Hughes, D. H., et al. 1998, Nature, 394, 241
\reference{} Ilbert, O., et al. 2009, ApJ, 690, 1236
\reference{} Kauffmann, G., et al. 2003, MNRAS, 341, 54
\reference{} Kennicutt, R. C. 1998, ARA\&A, 36, 189
\reference{} Keres, D., Katz, N., Weinberg, D. H.,\& Dav\'{e}, R. 2005, MNRAS, 363, 2
\reference{} Keres, D., Katz, N., Fardal, M., Dav\'{e}, R.,\& Weinberg, D. H. 2009, MNRAS, 395, 160
\reference{} Kewley, L. J.,\& Ellison, S. L. 2008, ApJ, 681, 1183
\reference{} Khochfar, S.,\& Silk, J. 2006, ApJ, 648, 21
\reference{} Koekemoer, A. M., Fruchter, A. S., Hook, R. N.,\& Hack, W. 2002, HST Calibration Workshop (eds. S. Arribas, A. M. Koekemoer, B. Whitmore; STScI: Baltimore), 337
\reference{} Koekemoer, A. M., et al. 2007, ApJS, 172, 196
\reference{} Koekemoer, A. M., et al. 2011, submitted to ApJS (arXiv1105.3754)
\reference{} Kriek, M., et al. 2006, ApJ, 649, 71
\reference{} Kriek, M., van Dokkum, P. G., Labb\'{e}, I., Franx, M., Illingworth, G. D., Marchesini, D.,\& Quadri, R. F. 2009a, ApJ, 700, 221
\reference{} Kriek, M., van Dokkum, P. G., Franx, M., Illingworth, G. D.,\& Magee, D. K. 2009b, ApJ, 705, 71
\reference{} Krumholz, M. R., McKee, C. F.,\& Tumlinson, J. 2009, ApJ, 693, 216
\reference{} Le Floc'h, E., et al. 2009, ApJ, 703, 222
\reference{} Lilly, S. J., et al. 2009, ApJS, 184, 218
\reference{} Luo, B., et al. 2008, ApJS, 179, 19
\reference{} Lutz, D., et al. 2011, A\&A, 532, 90
\reference{} Magnelli, B., Elbaz, D., Chary, R. R., Dickinson, M., Le Borgne, D., Frayer, D. T.,\& Willmer, C. N. A. 2009, A\&A, 496, 57
\reference{} Maier, C., et al. 2009, ApJ, 694, 1099
\reference{} Maiolino, R., et al. 2008, A\&A, 488, 463
\reference{} Mannucci, F., et al. 2009, MNRAS, 398, 1915
\reference{} Mannucci, F., Cresci, G., Maiolino, R., Marconi, A.,\& Gnerucci, A. 2010, MNRAS, 408, 2115
\reference{} Maraston, C., Pforr, J., Renzini, A., Daddi, E., Dickinson, M., Cimatti, A.,\& Tonini, C. 2010, MNRAS, 407, 830
\reference{} Martig, M., Bournaud, F., Teyssier, R.,\& Dekel, A. 2009, ApJ, 707, 250
\reference{} McGrath, E. J., Stockton, A., Canalizo, G., Iye, M.,\& Maihara, T. 2008, ApJ, 682, 303
\reference{} McLeod, K. K., Rieke, G. H., Rieke, M. J.,\& Kelly, D. M. 1993, ApJ, 412, 111
\reference{} Mihos, J. C.,\& Hernquist, L. 1996, ApJ, 464, 641
\reference{} Nagy, S. R., Law, D. R., Shapley, A. E.,\& Steidel, C. C. 2011, ApJL, 735, 19
\reference{} Noeske, K. G., et al. 2007, ApJ, 660, 43
\reference{} Nordon, R., et al. 2010, A\&A, 518, 24
\reference{} Peng, C. Y., Ho, L. C., Impey, C. D.,\& Rix, H.-W. 2010, AJ, 139, 2097
\reference{} Peng, Y. J., et al. 2010, ApJ, 721, 193
\reference{} Reddy, N. A., Erb, D. K., Pettini, M., Steidel, C. C., Shapley, A. E. 2010, ApJ, 712, 1070
\reference{} Rodighiero, G., et al. 2010, A\&A, 518, 25
\reference{} Rodighiero, G., et al. 2011, ApJL, 739, 40
\reference{} Salim, S., et al. 2007, ApJS, 173, 267
\reference{} Sanders, D. B., Soifer, B. T., Elias, J. H., Madore, B. F., Matthews, K., Neugebauer, G.,\& Scoville, N. Z. 1988, ApJ, 325, 74
\reference{} Sanders, D. B., et al. 2007, ApJS, 172, 86
\reference{} Sargent, M. T., et al. 2007, ApJS, 172, 434
\reference{} Scarlata, C., et al. 2007, ApJS, 172, 494
\reference{} Schiminovich, D., et al. 2007, ApJS, 173, 315
\reference{} Scoville, N. Z., et al. 2007, ApJS, 172, 1
\reference{} Shapiro, K. L., et al. 2008, ApJ, 682, 231
\reference{} Shen, S., Mo, H. J., White, S. D. M., Blanton, M. R., Kauffmann, G., Voges, W., Brinkmann, J.,\& Csabai, I. 2003, MNRAS, 343, 978
\reference{} Simmons, B. D.,\& Urry, C. M. 2008, ApJ, 683, 644
\reference{} Smail, I., Ivison, R. J.,\& Blain, A. W. 1997, ApJ, 490, 5
\reference{} Snyder, G. F., Cox, T. J., Hayward, C. C., Hernquist, L.,\& Jonsson, P. 2011, submitted to ApJ (arXiv1102.3689)
\reference{} Szomoru, D., Franx, M., Bouwens, R. J., van Dokkum, P. G., Labb\'{e}, I., Illingworth, G. D.,\& Trenti, M. 2011, ApJ, 735, 22
\reference{} Tacconi, L. J., et al. 2008, ApJ, 680, 246
\reference{} Tacconi, L. J., et al. 2010, Nature, 463, 781
\reference{} Toft, S., Franx, M., van Dokkum, P., F\"{o}rster Schreiber, N. M., Labb\'{e}, I., Wuyts, S.,\& Marchesini, D. 2009, ApJ, 705, 255
\reference{} Toomre, A.,\& Toomre, J. 1972, ApJ, 178, 623
\reference{} van der Wel, A., Holden, B. P., Zirm, A. W., Franx, M., Rettura, A., Illingworth, G. D.,\& Ford, H. C. 2008, ApJ, 688, 48
\reference{} Trujillo, I., et al. 2006, ApJ, 650, 18
\reference{} Ueda, Y., et al. 2008, ApJS, 179, 124
\reference{} van der Wel, A., et al. 2011, ApJ, 730, 38
\reference{} van Dokkum, P. G., et al. 2008, ApJ, 677, 5
\reference{} van Dokkum, P. G., et al. 2011, submitted to ApJL (arXiv1108.6060)
\reference{} Vanzella, E., et al. 2008, A\&A, 478, 83
\reference{} Williams, R. J., Quadri, R. F., Franx, M., van Dokkum, P. G., Toft, S., Kriek, M.,\& Labb\'{e}, I. 2010, ApJ, 713, 738
\reference{} Windhorst, R. A., et al. 2011, ApJS, 193, 27
\reference{} Wuyts, S., et al. 2007, ApJ, 655, 51
\reference{} Wuyts, S., Labb\'{e}, I., F\"{o}rster Schreiber, N. M., Franx, M., Rudnick, G., Brammer, G. B.,\& van Dokkum, P. G. 2008, ApJ, 682, 985
\reference{} Wuyts, S., Franx, M., Cox, T. J., Hernquist, L., Hopkins, P. F., Robertson, B. E.,\& van Dokkum, P. G. 2009, ApJ, 696, 348
\reference{} Wuyts, S., Cox, T. J., Hayward, C. C., Franx, M., Hernquist, L., Hopkins, P. F., Jonsson, P.,\& van Dokkum, P. G. 2010, ApJ, 722, 1666
\reference{} Wuyts, S., et al. 2011, ApJ, 738, 106

}
\end {references}



\end {document}